\documentclass[11pt]{article}
\usepackage[letterpaper,margin=1.2in]{geometry}

\usepackage{setspace}
\usepackage{etoolbox}

\AtBeginEnvironment{thebibliography}{%
  \setstretch{1}\normalsize
}

\usepackage{graphicx}
\usepackage{tabularx}
\usepackage{array}
\usepackage{amsmath}
\usepackage{amsthm}
\usepackage{booktabs}
\usepackage{float}
\usepackage{placeins}
\usepackage{pifont}
\usepackage{multirow}
\usepackage{longtable}
\usepackage{adjustbox}
\usepackage{pdflscape}
\usepackage[normalem]{ulem}
\usepackage{amssymb}
\usepackage{amsfonts}
\usepackage[numbers,square,sort&compress]{natbib}

\usepackage{lineno}

\usepackage[colorlinks=true,linkcolor=blue,citecolor=blue,urlcolor=blue]{hyperref}

\newcommand{\cmark}{\ding{51}}

\title{From GEV to ResLogit: Spatially Correlated Discrete Choice Models for Pedestrian Movement Prediction}

\author{%
Rulla Al-Haideri\textsuperscript{1}\thanks{Corresponding author. Email: rullaalhaideri@torontomu.ca} \and
Bilal Farooq\textsuperscript{1}\thanks{Email: bilal.farooq@torontomu.ca}
}

\date{}

\begin{document}
\onehalfspacing
\maketitle

\begin{center}
\textsuperscript{1}Laboratory of Innovations in Transportation (LiTrans), Toronto Metropolitan University, 350 Victoria Street, Toronto, Ontario, Canada
\end{center}

\begin{abstract}
\noindent
High frequency pedestrian motion forecasting when interacting with autonomous vehicles (AVs) can be enhanced through the use of behavioural frameworks, such as discrete choice models, that can explicitly account for correlation among similar movement alternatives. We formulate the pedestrian next step choice as a spatial discrete choice defined by a grid of speed adjustment and heading change. 
Using naturalistic pedestrian--AV encounters from nuScenes and Argoverse~2 (1~sec decision interval), we estimate a multinomial logit baseline and four spatial generalized extreme value (GEV) specifications (SCL, GSCL, SCNL, and GSCNL). We then compare them to a residual neural network logit (ResLogit) model that learns cross alternative effects while retaining an interpretable linear utility component. 
Across the evaluated data, spatial GEV structures yield only marginal improvements over multinomial logit, whereas ResLogit achieves a substantially better fit and produces behaviourally coherent errors concentrated among neighbouring grid cells. 
The results suggest that in dense, high frequency spatial choice sets, learning based residual corrections can capture proximity induced correlation more effectively than analyst specified GEV nesting structures, while maintaining interpretability.

\end{abstract}


\section{Introduction}
In geographic and spatio-temporal modelling, distance is a foundational determinant of human motion patterns. Tobler's first law states that nearby entities tend to be more related than distant ones \cite{Tobler1970}. In discrete choice modelling, this implies that alternatives located close to each other can share unobserved attributes, which induces correlation among their utilities. This is a particular concern in spatial choice settings, where neighbouring alternatives are often close substitutes for reasons the analyst cannot fully observe.

A substantial literature focuses on modelling spatially correlated alternatives, particularly in location choice. 
Early approaches rely on flexible error structures, most commonly within probit and mixed logit frameworks. 
Spatial autoregressive multinomial probit formulations introduce correlation between geographically proximate alternatives through exogenously specified spatial weight matrices \cite{BolducFortinGordon1997}. 
Spatial autoregressive processes were later embedded in both the deterministic and stochastic components of utility within simulation based mixed logit specifications \cite{MiyamotoEtAl2004}. 
Other related extensions introduce distance based parameterizations of the variance--covariance structure so that closer alternatives exhibit stronger correlation, while retaining a mixed logit kernel and simulation based likelihood evaluation \cite{WeissHabib2017}. These methods are expressive but they typically require non-trivial matrix constructions. They also require careful prior specification of spatial weighting schemes, and additional computational burden due to simulation.

In parallel, closed-form models within the generalized extreme value (GEV) family provide analytically convenient ways to relax independence while preserving tractable choice probabilities \cite{Train2009}. 
Spatial GEV formulations adapt a GEV generating function to capture correlation structures that reflect spatial relationships among alternatives. 
The spatially correlated logit (SCL) model relies on a contiguity matrix to allocate alternatives to overlapping nests under a single correlation parameter \cite{BhatGuo2004}. 
The generalized SCL (GSCL) replaces binary contiguity with a distance based allocation rule so that correlation decays with separation while retaining a single correlation parameter \cite{Sener2011}. The spatially correlated nested logit (SCNL) allows multiple nesting coefficients but retains analyst specified spatial linkage through contiguity based allocation \cite{PerezLopez2022}. 
Building on this, the generalized spatially correlated nested logit (GSCNL) aims to reduce reliance on predefined contiguity or distance decay assumptions by estimating allocation patterns from the data. This allows correlation strength and allocation to adapt to observed choice behaviour \cite{AlHaideriGSCNL}. 
These closed-form spatial GEV models are still attractive because they preserve interpretability and computational efficiency. However, these models still require the analyst to impose or justify a spatial linkage design which can become restrictive when choice sets are dense and decisions occur at high frequency.

This challenge becomes critical in high frequency pedestrian movement modelling. 
Next step walking decisions are inherently spatial because each decision selects one movement alternative from a spatially structured choice set at each decision instant. In our setting, alternatives are formed as a $3\times3$ grid defined by speed adjustment and heading change. 
Neighbouring movement alternatives are similar by construction. This makes proximity induced correlation intrinsic to the choice set. 
Predicting pedestrian motion under midblock interactions is also operationally important for autonomous vehicles (AVs). 
Planning and yielding decisions depend on short horizon predictions of how pedestrians adjust speed and heading during interaction pressure. 
A next step movement model supports pedestrian microsimulation at midblocks and provides a behavioural baseline for motion prediction. 
In subsequent work, this baseline will be extended to multi-step forecasting with temporal dependence and a richer scene and interaction context, while preserving the discrete choice structure and interpretability.

At the same time, pedestrian motion prediction in the vicinty of AVs has been dominated by machine learning approaches that predict future coordinates from past trajectories and scene context. Pedestrian trajectory prediction has been extensively studied using sequence modelling and representation learning. Early deep learning approaches used recurrent networks with social interaction modules such as social pooling are adopted to capture pedestrian--pedestrian interactions in crowded environments \cite{Alahi2016SocialLSTM,Gupta2018SocialGAN}. 
Subsequent work expanded to multimodal forecasting, graph structured interaction modelling, and map aware prediction such as modular graph recurrent architectures and vectorized map encoders \cite{Ivanovic2019Trajectron,Salzmann2020TrajectronPlusPlus,Gao2020VectorNet}. 
Many methods included scene context, distance to vehicles, head orientation, or other behavioural cues to improve prediction at crossings and midblock interactions \cite{KalatianFarooq2022ContextAware}. 
Related research have also focused on pedestrian crossing intention prediction and behaviour understanding. They framed the task as intention classification or joint intention and trajectory prediction \cite{RasouliTsotsos2019Survey,Landry2024CrossingIntentionReview}.

Despite these advances, two gaps are directly relevant. 
First, many trajectory prediction models focused on forecasting accuracy and powerful feature learning. However, they did not produce a behavioural model that quantified how inputs change choice probabilities (i.e., elasticities) or how probability shifts between neighbouring movement alternatives within a structured choice set. 
Second, in decision making applications, the most important prediction mistakes are often those that stay within the same small neighbourhood of actions. For a $3\times3$ movement grid, predicting an adjacent cell (i.e., a slight change in speed adjustment or heading) is typically much less harmful than predicting a distant cell that represents a qualitatively different action (e.g., predicting acceleration when the pedestrian actually decelerated). 
These reasons motivate modelling pedestrian motion as a spatial choice over a discretized movement grid and evaluating models based on whether their errors and correlation structures remain behaviourally coherent within the local neighbourhood of feasible movements.
For these reasons, we formulate pedestrian next step motion as a discrete choice problem in this work. 
At each decision instant, the pedestrian selects one alternative from a movement grid defined by speed adjustment and heading change. 
The discrete choice formulation provided a behavioural decision rule with structurally interpretable parameters in the linear utility. This enables direct statements about trade offs and the directional influence of interaction variables, while producing explicit substitution patterns across nearby alternatives.

Recent hybrid choice models are proposed to reduce reliance on analyst defined correlation structures by learning systematic cross effects directly from data. Wong and Farooq introduce the residual neural network logit (ResLogit) model. This model augments a multinomial logit (MNL) backbone with residual layers that learn non-linear cross effects and unobserved behavioural structure, while still supporting behavioural indicators such as substitution patterns and elasticities \cite{WongFarooq2021}. 
For ordered responses, Kamal and Farooq extend this approach to Ordinal-ResLogit by integrating ResLogit with a consistent rank logits construction \cite{KamalFarooq2024}. These models represent data driven approaches to capture correlation and heterogeneity that would could be forced into analyst defined nests, contiguity matrices, or distance decay specifications.

This paper frames next step of pedestrian movement choice as a spatial choice problem. It evaluates whether classic spatial GEV models remain adequate for high frequency movement choices, or whether hybrid learning based choice models provide a better representation of proximity induced correlation. Naturalistic midblock pedestrian crossing decisions with AV interactions are extracted from two publicly available trajectory datasets, nuScenes \cite{caesar2020nuscenes} and Argoverse 2 \cite{Wilson2021Argoverse2}. The main objective is to compare an MNL and spatial GEV formulations (SCL, GSCL, SCNL, and GSCNL) with ResLogit in terms of their ability to capture spatial correlation in the movement choice grid and their predictive performance.

The rest of the paper is organized as follows. 
Section~2 describes the datasets, the spatial choice set construction, and the model formulations including spatial GEV and ResLogit. 
Section~3 presents estimation results, prediction performance, and diagnostic analyses to interpret why classic spatial correlation structures provide limited gains in dense movement grids. 
Section~4 concludes with implications for behaviourally grounded motion prediction and highlights extensions for sequential and multi-agent modelling.

\section{Literature Review}

\subsection{Pedestrian motion prediction}

Pedestrian motion prediction has long been treated as a core perception to planning capability for AVs. This is because short horizon uncertainty directly affects yielding, collision avoidance, and comfort constrained manoeuvres. 
Early data driven work established sequence-to-sequence pedestrian trajectory prediction using recurrent neural networks and introduced explicit interaction modelling through social pooling to capture neighbouring trajectories in crowds \cite{Alahi2016SocialLSTM}.
Social-GAN extended this idea to a generative framework that produced diverse and socially plausible futures which focuse on multimodal and interaction consistency rather than a single best prediction \cite{Gupta2018SocialGAN}.

As trajectory prediction matured in AV settings, methods captured heterogeneous scene context increasingly. These include semantic maps, lane geometry, and agent dynamics, and interaction structure for multiple road users. 
Trajectron++ proposed a graph structured recurrent architecture that handled heterogeneous agent types and included environmental information while supporting multimodal predictions \cite{Salzmann2020TrajectronPlusPlus}. 
VectorNet introduced a vectorized representation of map elements and trajectories\cite{Gao2020VectorNet}. This framework used a hierarchical graph network to capture spatial locality and global interactions. It demonstrated competitive performance on large-scale autonomous driving benchmarks. 
Such approaches reflect a dominant paradigm in which pedestrian motion is forecast as continuous future coordinates linked to past motion and scene context \cite{Salzmann2020TrajectronPlusPlus,Gao2020VectorNet}.
In contrast, our study formulates short horizon pedestrian motion as a discrete choice among structured movement alternatives. 
This enables interpretable trade-offs and explicit substitution patterns across alternatives.
This formulation therefore yields a calibrated probability distribution over candidate movement alternatives. This formulation can be directly used in hypothesis based AV planning, and evaluated for local coherence in substitution and misclassification.

Two limitations of the trajectory forecasting literature are directly relevant to behaviourally grounded motion prediction. 
First, many high capacity predictors focus on representating learning and predictive accuracy but do not provide an explicit behavioural decision rule that supports interpretable trade-offs, elasticities, or substitution patterns across discrete movement alternatives. 
Second, with regards to control and risk assessment, not all prediction errors are equally consequential. 
Confusing two neighbouring movement alternatives often has little operational impact. Confusing qualitatively different alternatives can lead to substantially different risk and planning decisions.
This motivates the evaluation and modelling designs that focus on local coherence in predicted probability distribution and misclassification structure. This is particular when the prediction problem is defined over a structured set of candidate actions instead of continuous coordinates.

\subsection{Structured action spaces and discretization in motion forecasting}

A substantial body of litarture adopted discretized or structured action sets to represent multimodality and to enable tractable probabilistic reasoning over future behaviours. 
In AVs, classification based forecasting over a finite set of candidate trajectories has been used to manage the intrinsic multimodality of human motion while keeping inference efficient. 
CoverNet presented motion prediction as classification by a curated set of feasible trajectories \cite{PhanMinh2020CoverNet}. The authors argued that the number of distinct actions over a practical horizon is limited and that a discrete trajectory set can capture diverse futures. 
MultiPath used a fixed set of anchor trajectories representing modes of the future distribution and predicted a discrete distribution over anchors with continuous offsets \cite{Chai2020MultiPath}. This yielded a compact mixture representation suitable for probabilistic queries. 
Related work has also treated trajectory prediction as ranking over a trajectory bank where the model assigns higher scores to more plausible candidates \cite{Biktairov2020PRANK}.

These designs formalize an important point for pedestrian forecasting in AV contexts which is that multimodality is often better represented as selection among competing motion hypotheses than as a single regressed path. 
Discretization provides a natural interface between prediction and planning. Planners frequently evaluate a finite set of candidate pedestrian motions when assessing collision risk and safe responses. 
Nevertheless, discretization introduces technical challenge that must be addressed in a behavioural modelling context. 
Discrete action sets can be dense and symmetric and class frequencies can be imbalanced. In addition, nearby alternatives can be near substitutes whose utilities share unobserved components. 
These properties tend to amplify correlation across alternatives and can weaken identifiability of correlation structures that rely on strong analyst imposed geometry. 
Hence, when the choice set is defined as a structured grid of micro-actions, the key modelling question shifts from whether multimodality exists into how local similarity and cross alternative correlation should be captured so that probability distributions and errors remain locally coherent.

\subsection{Correlation across alternatives in discrete choice}

In random utility models, the MNL remains the most widely used baseline because it is closed-form and interpretable. 
Its main limitation is the independence of irrelevant alternatives property, which is implied by i.i.d.\ type-I extreme value errors \cite{Train2009}. 
In dense and structured choice sets, independence can become unrealistic since alternatives that are close in some behavioural sense. An example to that is spatial proximity or similarity in speed and heading change. 
Such set up of alternatives often share unobserved attributes. 
When such correlation is present, an MNL can misrepresent substitution patterns by shifting probability distributions too aggressively across dissimilar alternatives.

Two broad families of approaches have been used to relax independence. Simulation-based or open-form models (e.g., multinomial probit and mixed logit) which can accommodate flexible error correlation and heterogeneity. 
However, these models typically require nontrivial covariance specifications and simulation based likelihood evaluation. This can be computationally expensive for repeated estimation and for large or complex choice sets \cite{Train2009}. 
Closed-form models within the GEV family provide an alternative to correlated utilities while preserving tractability through an analytically defined generating function \cite{Train2009}. 
Classic examples include nested logit and more general cross-nested structures. These which can induce correlation among subsets of alternatives. 
The practical challenge is that closed-form correlation requires the analyst to define, justify, and operationalize the similarity structure that determines how alternatives share nests. Also, the design choice can dominate results when the choice set is dense and the similarity geometry is ambiguous.
These issues become particularly critical for discretized motion choice sets. 
When alternatives differ only by small micro-adjustments, correlation is intrinsic. However, the correct correlation topology is rarely obvious. 
In such settings, the evaluation target is not only improved fit but also to become behaviourally coherent substitution. 
In other words, the probability distributions and misclassifications should concentrate among nearby alternatives that represent plausible micro-variants of the observed action. 
Therefore, this provides a direct motivation for spatial and metric correlation models, and for learning based approaches that infer cross alternative structure from data.

\subsection{Discrete choice models for pedestrian motion and correlated alternatives}

Discrete choice models have been used extensively in transportation to represent decision making and to obtain interpretable behavioural parameters. 
In pedestrian dynamics, this framework has been applied directly to short horizon walking decisions by modelling the next step as a choice among spatially defined movement alternatives. Antonini et al. proposed a discrete choice formulation for pedestrian walking dynamics with a spatial discretization of the environment and estimated a cross nested and mixed nested logit models using video tracked trajectories \cite{AntoniniBierlaireWeber2006}. 
Robin et al. further advanced this line of research by specifying, estimating, and validating a pedestrian walking behaviour model that distinguished unconstrained from interaction constrained behaviour \cite{RobinAntoniniBierlaireCruz2009}. They captured spatial correlation using a cross nested logit structure. 
In the context of road crossing, discrete choice models has also been used to represent microscopic cyclist and pedestrian actions under vehicle induced hazards and heterogeneity, for example through an MNL and latent class specifications for crossing decisions at roundabout crosswalks \cite{AlHaideriWeissIsmailRoundaboutUnderReview, alhaideri2025cyclistAAP}. 
More recently, hybrid learning based discrete choice models has been used to model pedestrian movement adjustments during AV encounters in naturalistic data, while retaining utility based interpretability \cite{AlHaideriFarooq2026PedAV}.

A key challenge in next step motion choice is how to represent correlation among spatially or behaviourally similar alternatives. 
In dense and structured action grids, neighbouring alternatives are near substitutes by construction and may share unobserved attributes. This could violates the independence assumptions of the MNL \cite{Train2009}. 
Closed-form models within the GEV family address this issue by inducing correlation through structured error components, while preserving tractable choice probabilities. 
Spatial GEV formulations operationalize similarity by allocating alternatives to overlapping nests using contiguity or distance based linkage rules \cite{BhatGuo2004,Sener2011,PerezLopez2022,AlHaideriGSCNL}.
These models are attractive for high frequency motion choice because they preserve interpretability and analytical probabilities.
However, their performance depends on whether the analyst specified linkage design matches the true local substitution patterns present in the data. 
When the choice set is small, symmetric, and densely packed, linkage geometry can be weakly identified and overlapping nests can distribute correlation across many components. 
This could reduce the practical impact of nesting parameters on predicted probability allocation.

Learning based discrete choice hybrids reduce reliance on analyst imposed correlation structure by learning cross alternative effects directly from data while retaining a behavioural utility backbone. 
ResLogit augments a logit model with residual layers that learn non-linear cross effects and unobserved structure, while preserving an interpretable linear utility component \cite{WongFarooq2021ResLogit}. 
In structured motion grids, similarity is local but context dependent. The residual learning can capture systematic probability re-allocation among neighbouring actions without discarding interpretability. 
Therefore, comparing spatial GEV models to ResLogit in a discretized pedestrian motion choice setting provides a direct test of whether analyst imposed metric correlation remains adequate at high frequency, or whether learned residual corrections better represent proximity induced correlation in dense action spaces.

\subsection{Research gap}

Existing work on pedestrian motion prediction and pedestrian walking behaviour leaves a gap at the intersection of behavioural interpretability, spatial correlation modelling, and high frequency action discretization.
First, the trajectory forecasting literature for AV contexts has largely focused on continuous coordinate prediction using high capacity sequence and graph based neural architectures. 
These models achieve strong predictive performance, they typically do not provide an explicit behavioural decision rule, nor do they characterize substitution patterns across structured movement alternatives. 
Therefore, it is difficult to interpret how risk, proximity, or goal orientation trade-offs shift the relative likelihood of competing micro-actions.
Second, discrete choice models have been successfully applied to pedestrian walking and crossing behaviour. This includes short term step decisions and interaction aware movement adjustments. 
However, most existing applications either rely on MNL specifications that impose independence across alternatives, or spatial cross nesting structures that are designed for larger spatial choice sets such as location choice. The behaviour of these correlation structures in dense, symmetric, and small action grids remains insufficiently examined.
Third, spatial GEV formulations introduce correlation through analyst specified contiguity or distance decay rules. 
These linkage designs may become weakly identified when alternatives are highly similar by construction, although analytically convenient. This can be the case in discretized motion grids. 
In such settings, it is unclear if predefined spatial nesting structures meaningfully affect substitution patterns beyond the MNL baseline.
Finally, recent learning based hybrid discrete choice models offer a data driven approach to capture correlation in cross alternatives while preserving an interpretable utility backbone. 
Yet, their comparative performance relative to spatial GEV models has not been systematically evaluated in high frequency pedestrian motion choice settings with structured action spaces.
Therefore, we define a clear methodological gap. It remains unknown whether analyst imposed spatial correlation structures are sufficient to represent proximity induced correlation in dense pedestrian movement grids, or whether residual learning based corrections provide a more behaviourally coherent representation of local substitution. 
This paper addresses this gap by formulating pedestrian next step motion choice as a spatial discrete choice problem and directly comparing an MNL, spatial GEV models (SCL, GSCL, SCNL, and GSCNL), and ResLogit within the same discretized movement framework.

\section{Methods}
\begin{figure}[H]
\centering
\includegraphics[width=0.5\linewidth]{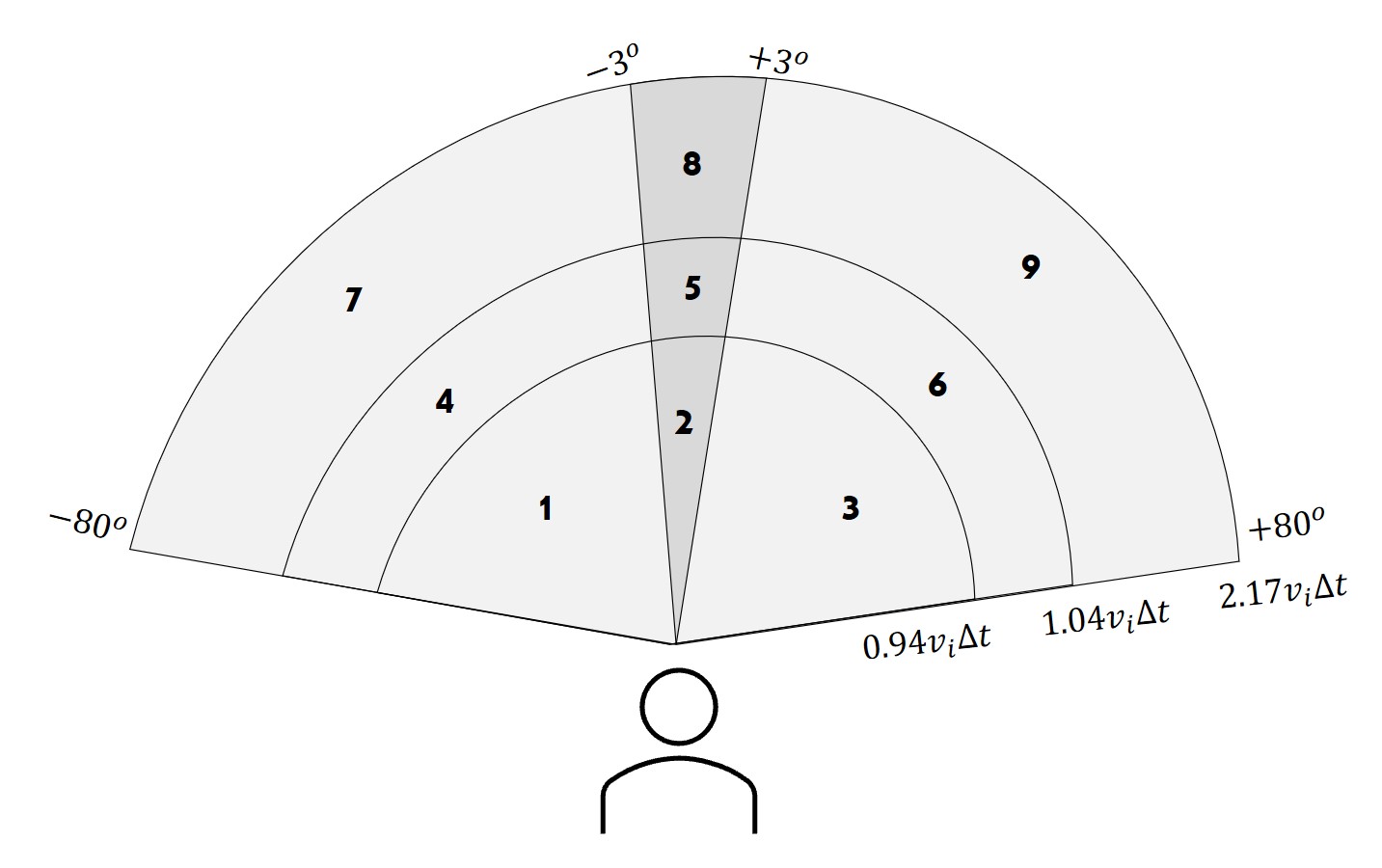}
\caption{Pedestrian spatial choice set grid.}
\label{fig:spatial_grid}
\end{figure}

This study presents a discrete choice model for pedestrian motion prediction. It is designed to isolate how alternative correlation behaves in a dense $3\times3$ movement grid (Figure~\ref{fig:spatial_grid}). Extensions to multi-step forecasting with temporal dependence and richer interaction context are reserved for subsequent work.
\subsection{Dataset}
Two publicly available trajectory datasets including nuScenes \cite{caesar2020nuscenes} and Argoverse 2 \cite{Wilson2021Argoverse2} are used that capture naturalistic pedestrian interactions with an ego AV at 1~sec decision interval. 
Building on our prior ResLogit framework where we modelled pedestrian speed adjustment using the nuScenes dataset and by defining acceleration and deceleration alternatives (two choices) \cite{AlHaideriFarooq2026PedAV}, we extend the setting in two ways. 
First, we include the Argoverse 2 dataset in addition to the nuScenes. Second, we expand the next step choice set from speed only adjustments to a $3\times3$ movement grid that jointly represents speed adjustment (decelerate, maintain speed, accelerate) and heading change (left, maintain direction, right). Each choice represents the potential movement position (i.e., the choice) at the next time step of the pedestrian. Figure~\ref{fig:spatial_grid} illustrates the spatial movement choice set grid.
Each observed choice represents the pedestrian's next step movement position and is characterized by two variables computed at $t+1$: the relative change in speed and the relative change in heading. These two variables jointly map the observed positions onto one of the nine cells in the spatial grid.
The grid boundaries are defined empirically based on the observed distribution of one step speed and heading changes in the trajectory data. Additional details on the choice set boundary construction for a similar framework are provided in  \cite{alhaideri2025cyclistAAP,alhaideri2025cttc}. 
Some steps cannot be assigned a valid label. This occurs when the implied speed change ratio or heading change falls outside the pre-defined grid boundaries, or when the observation corresponds to the final time step of a pedestrian trajectory and no next step exists. These steps are excluded from model estimation. Table~\ref{tab:choice_label_counts} summarizes the total number of extracted decision steps, the number excluded, and the distribution of valid labels.
For Argoverse 2, 1{,}133 decision steps (i.e., valid observed choices) are extracted, of which 863 are retained after filtering. 
For nuScenes, 2{,}170 decision steps are extracted, of which 1{,}830 are retained. 
For all retained steps, kinematic, spatial, and interaction variables are computed at each decision instant and used as inputs to the discrete choice models.

\begin{table}
\centering
\caption{Observed movement choice counts before and after filtering to the defined $3\times3$ grid.}
\label{tab:choice_label_counts}
\small
\begin{tabularx}{0.7\linewidth}{>{\raggedright\arraybackslash}X rr}
\toprule
 & Argoverse 2 & nuScenes \\
\midrule
Total decision steps (raw) & 1133 & 2170 \\
Excluded steps (outside grid or final step) & 270 & 340 \\
Processed decision steps (valid 1--9) & 863 & 1830 \\
\midrule
Choice 1 & 65 & 180 \\
Choice 2 & 55 & 222 \\
Choice 3 & 92 & 136 \\
Choice 4 & 81 & 150 \\
Choice 5 & 144 & 398 \\
Choice 6 & 109 & 150 \\
Choice 7 & 110 & 155 \\
Choice 8 & 106 & 266 \\
Choice 9 & 101 & 173 \\
\bottomrule
\end{tabularx}
\end{table}

\subsection{Model Structure}
Let $t\in\{1,\ldots,N\}$ index decision steps (i.e., time steps) and let $i\in\mathcal{C}=\{1,\ldots,J\}$ index the $J=9$ movement alternatives in the $3\times3$ grid. The total utility of alternative $i$ at decision step $t$ is given by
\begin{equation}
U_{it}=V_{it}+\varepsilon_{it},
\end{equation}
where $V_{it}$ is the systematic component and $\varepsilon_{it}$ captures unobserved influences. Under the MNL model, $\varepsilon_{it}$ are i.i.d.\ type-I extreme value. The choice probability is expressed as
\begin{equation}
P^{\text{MNL}}_{it}=\frac{e^{V_{it}}}{\sum_{j\in\mathcal{C}} e^{V_{jt}}}.
\label{eq:mnl_prob}
\end{equation}

To relax independence for spatial alternatives, we consider the spatial GEV family. The joint distribution of the error terms is generated by a GEV generating function $G(\cdot)$ \cite{Train2009}. Following the SCL model \cite{BhatGuo2004}, correlation is represented through overlapping pairwise nests. For each unordered pair $(i,m)$ with $i<m$, a nest $(im)$ is defined with allocation parameters $\alpha_{i,im}$ and $\alpha_{m,im}$ satisfying $0<\alpha_{i,im}<1$ and $\sum_{m\neq i}\alpha_{i,im}=1$ for each $i$. The SCL generating function is given by
\begin{equation}
G\!\left(e^{V_{1t}},\ldots,e^{V_{Jt}}\right) = \sum_{i=1}^{J-1}\sum_{m=i+1}^{J}
\Big[ 
\left(\alpha_{i,im} e^{V_{it}}\right)^{1/\lambda}
+
\left(\alpha_{m,im} e^{V_{mt}}\right)^{1/\lambda}
\Big]^{\lambda},
\label{eq:scl_generator}
\end{equation}
where $\lambda\in(0,1]$ is the nesting coefficient, and when $\lambda=1$, the model collapses back to an MNL.

Given $G(\cdot)$, the choice probability under the GEV class is expressed as
\begin{equation}
P_{it}
=
\frac{
e^{V_{it}}\;\dfrac{\partial G}{\partial e^{V_{it}}}
}{
G\!\left(e^{V_{1t}},\ldots,e^{V_{Jt}}\right)
}.
\label{eq:gev_prob_general}
\end{equation}
For the SCL generator in Equation~\eqref{eq:scl_generator}, the probability can be expressed as a sum over pairwise nests \cite{BhatGuo2004}:
\begin{equation}
P^{\text{SCL}}_{it}=\sum_{m\neq i} P_{t}(i\mid im)\,P_{t}(im).
\label{eq:scl_decomp}
\end{equation}
The conditional probability of choosing $i$ given nest $(im)$ is
\begin{equation}
P_{t}(i\mid im)=
\frac{\left(\alpha_{i,im} e^{V_{it}}\right)^{1/\lambda}}
{\left(\alpha_{i,im} e^{V_{it}}\right)^{1/\lambda}+\left(\alpha_{m,im} e^{V_{mt}}\right)^{1/\lambda}}.
\label{eq:scl_cond}
\end{equation}
We define the pairwise inclusive value as
\begin{equation}
S_{im,t}=
\left(\alpha_{i,im} e^{V_{it}}\right)^{1/\lambda}
+\left(\alpha_{m,im} e^{V_{mt}}\right)^{1/\lambda}.
\label{eq:scl_S}
\end{equation}
Then the probability of selecting nest $(im)$ is
\begin{equation}
P_{t}(im)=
\frac{S_{im,t}^{\lambda}}{\sum_{s=1}^{J-1}\sum_{l=s+1}^{J} S_{sl,t}^{\lambda}}.
\label{eq:scl_nestprob}
\end{equation}

In the SCL model, the allocation is defined using a contiguity indicator $\omega_{im}$ that equals one if alternatives $i$ and $m$ are spatially contiguous and zero otherwise \cite{BhatGuo2004}:
\begin{equation}
\alpha_{i,im}=\frac{\omega_{im}}{\sum_{s\neq i}\omega_{is}}.
\label{eq:scl_alpha}
\end{equation}
The GSCL model replaces contiguity with a distance based allocation rule so that correlation decays with separation \cite{Sener2011}:
\begin{equation}
\alpha_{i,im}=\frac{d_{im}^{\vartheta}}{\sum_{s\neq i}d_{is}^{\vartheta}},
\qquad \vartheta<0,
\label{eq:gscl_alpha}
\end{equation}
where $d_{im}$ is the distance between alternatives $i$ and $m$ in the spatial choice set.

The SCNL model allows multiple nesting coefficients across predefined nest categories \cite{PerezLopez2022}. Let $k\in\{1,\ldots,K\}$ index nest categories and let $\lambda_k$ denote the nesting coefficient for category $k$. For each pairwise nest $(im)$, define $k(im)$ as the category assigned to pair $(i,m)$ and set $\lambda_{im}\equiv \lambda_{k(im)}$. The choice probability is written as a sum over pairwise nests:
\begin{equation}
P^{\text{SCNL}}_{it}=\sum_{m\neq i} P^{\text{SCNL}}_{t}(i\mid im)\,P^{\text{SCNL}}_{t}(im).
\label{eq:scnl_decomp}
\end{equation}
The conditional probability of choosing $i$ given nest $(im)$ is
\begin{equation}
P^{\text{SCNL}}_{t}(i\mid im)=
\frac{\left(\alpha_{i,im} e^{V_{it}}\right)^{1/\lambda_{im}}}
{\left(\alpha_{i,im} e^{V_{it}}\right)^{1/\lambda_{im}}+\left(\alpha_{m,im} e^{V_{mt}}\right)^{1/\lambda_{im}}}.
\label{eq:scnl_cond}
\end{equation}
Let the pairwise inclusive value for SCNL be expressed as
\begin{equation}
S^{\text{SCNL}}_{im,t}=
\left(\alpha_{i,im} e^{V_{it}}\right)^{1/\lambda_{im}}
+\left(\alpha_{m,im} e^{V_{mt}}\right)^{1/\lambda_{im}}.
\label{eq:scnl_S}
\end{equation}
Then the nest probability is
\begin{equation}
P^{\text{SCNL}}_{t}(im)=
\frac{\left(S^{\text{SCNL}}_{im,t}\right)^{\lambda_{im}}}
{\sum_{s=1}^{J-1}\sum_{l=s+1}^{J} \left(S^{\text{SCNL}}_{sl,t}\right)^{\lambda_{sl}}}, 
\label{eq:scnl_nestprob}
\end{equation}
where $\lambda_{sl}\equiv \lambda_{k(sl)}$.

The GSCNL model further increases flexibility by estimating the allocation pattern from data rather than fixing it using a contiguity matrix or a distance-decay rule \cite{AlHaideriGSCNL}. The allocation parameters are specified using a pseudo-logit form over candidate nest memberships:
\begin{equation}
\alpha_{i,im} = \frac{ e^{\sum_{p}\delta_{p} I_{imp}}\,I_{im} }{ \sum_{q\neq i} e^{\sum_{p}\delta_{p} I_{iqp}}\,I_{iq} }.
\label{eq:gscnl_alpha}
\end{equation}
where $I_{imp}$ indicates whether the pair $(i,m)$ belongs to nesting category $p$, $I_{im}$ indicates whether $(i,m)$ forms a nest, and $\delta_p$ are parameters to be estimated which control allocation across categories. Nesting coefficients are category specific, $\lambda_k\in(0,1]$. The resulting choice probabilities follow the SCNL decomposition in Equations~\eqref{eq:scnl_decomp}--\eqref{eq:scnl_nestprob}, with allocation parameters given by Equation~\eqref{eq:gscnl_alpha}.

The ResLogit \cite{WongFarooq2021} first computes the MNL systematic utilities $\mathbf{V}_t=(V_{1t},\ldots,V_{Jt})^\top$, then applies a sequence of residual layers to obtain corrected utilities. We define $\mathbf{h}^{(0)}_t=\mathbf{V}_t$ and update
\begin{multline}
\mathbf{h}^{(m)}_t
=
\mathbf{h}^{(m-1)}_t
-
\mathrm{softplus}\!\left(\mathbf{h}^{(m-1)}_t \mathbf{W}^{(m)}\right),
m=1,\ldots,M,
\label{eq:reslogit_layers}
\end{multline}
where $\mathbf{W}^{(m)}\in\mathbb{R}^{J\times J}$ is a learnable weight matrix and $\mathrm{softplus}(x)=\ln(1+e^{x})$ is applied element-wise. The corrected observed utilities are $\mathbf{U}_t=\mathbf{h}^{(M)}_t$, and defining $\mathbf{g}_t=\mathbf{U}_t-\mathbf{V}_t$ yields $U_{it}=V_{it}+g_{it}$.
Assuming i.i.d.\ type-I extreme value errors around the corrected utilities, the implied choice probability is
\begin{equation}
P^{\text{ResLogit}}_{it}
=
\frac{e^{U_{it}}}{\sum_{j\in\mathcal{C}} e^{U_{jt}}}
=
\frac{e^{V_{it}+g_{it}}}{\sum_{j\in\mathcal{C}} e^{V_{jt}+g_{jt}}}.
\label{eq:reslogit_prob}
\end{equation}
When the residual component is identically zero, the model collapses back to an MNL.

\subsection{Models Specification}
Table~\ref{tab:model_variable_matrix} summarizes the components included in each model and highlights the differences between the closed-form spatial GEV formulations and the learning based ResLogit. 
Across all specifications, the systematic utility $V_{it}$ is constructed from the same set of movement and interaction variables (Table~\ref{tab:variable_definitions}). 
The models differ in how correlation and substitution across alternatives are represented. In spatial GEV models, it is captured through allocation parameters $\alpha_{i,im}$ and nesting coefficients $\lambda$ (or $\lambda_k$). In ResLogit, it is captured through a residual correction applied to the utility vector.

\begin{table*}
\centering
\caption{Model components included in each specification.}
\label{tab:model_variable_matrix}
\small
\setlength{\tabcolsep}{4pt}
\begin{tabular}{lcccccccccc}
\toprule
Model & $I^{\text{dec}}$ & $I^{\text{acc}}$ & $I^{\text{turn}}$ & FCRP & RCRP & $1/D$ & ddist & ddir & $\alpha$ & $\lambda$ \\
\midrule
ResLogit & \cmark & \cmark & \cmark & \cmark & \cmark & \cmark & \cmark & \cmark &  &  \\
MNL      & \cmark & \cmark & \cmark & \cmark & \cmark & \cmark &  &  &  &  \\
SCL      & \cmark & \cmark & \cmark & \cmark & \cmark & \cmark &  &  & \cmark & \cmark \\
GSCL     & \cmark & \cmark & \cmark & \cmark & \cmark & \cmark &  &  & \cmark$^{a}$ & \cmark \\
SCNL     & \cmark & \cmark & \cmark & \cmark & \cmark & \cmark &  &  & \cmark & \cmark$^{b}$ \\
GSCNL    & \cmark & \cmark & \cmark & \cmark & \cmark & \cmark &  &  & \cmark$^{c}$ & \cmark$^{b}$ \\
\bottomrule
\end{tabular}

\vspace{2pt}
\footnotesize
$^{a}$ GSCL uses distance based allocation (distance decay) in $\alpha_{i,im}$ in place of contiguity. \\
$^{b}$ SCNL and GSCNL allow multiple nesting coefficients $\lambda_k$ across predefined nest categories. \\
$^{c}$ GSCNL estimates allocation parameters $\alpha_{i,im}$ from data using a pseudo-logit form over candidate nest memberships.
\end{table*}

\begin{table}[t]
\centering
\caption{Definitions of model variables and parameters.}
\label{tab:variable_definitions}
\small
\setlength{\tabcolsep}{5pt}
\begin{tabularx}{\linewidth}{lX}
\toprule
Symbol & Definition \\
\midrule
$I^{\text{dec}}_{i}$ & Indicator for deceleration alternatives (grid row: decelerate). \\
$I^{\text{acc}}_{i}$ & Indicator for acceleration alternatives (grid row: accelerate). \\
$I^{\text{turn}}_{i}$ & Indicator for turning alternatives (left or right cones). \\
$D_t$ & AV--pedestrian distance at decision step $t$ (specified as $1/D_t$). \\
FCRP$_t$ & Front collision risk proximity at decision step $t$ (front intensity scaled by $1+\mathrm{CTTC}_t$). \\
RCRP$_t$ & Rear collision risk proximity at decision step $t$ (rear intensity scaled by $1+\mathrm{CTTC}_t$). \\
ddist$_{it}$ & Alternative-specific attraction-to-destination distance metric for alternative $i$ at step $t$. \\
ddir$_{it}$ & Alternative-specific attraction-to-destination direction metric for alternative $i$ at step $t$. \\
$\alpha_{i,im}$ & Allocation parameter assigning a share of alternative $i$ to pairwise nest $(im)$ in spatial GEV models. \\
$\lambda$ & Nesting coefficient controlling within-nest correlation (SCL, GSCL). \\
$\lambda_k$ & Category-specific nesting coefficient (SCNL, GSCNL). \\
\bottomrule
\end{tabularx}
\end{table}

The systematic utility adopts a parsimonious specification in the ResLogit model with eight parameters and predefined alternative mappings consistent with the $3\times3$ movement grid:
\begin{equation}
\begin{aligned}
V_{it}
=&\;
\beta_{\text{dec}} I^{\text{dec}}_{i}
+\beta_{\text{acc}} I^{\text{acc}}_{i}
+\beta_{\text{turn}} I^{\text{turn}}_{i}
+\beta_{\text{ddist}}\,\mathrm{ddist}_{it}
\\
&\;
+\beta_{\text{ddir}}\,\mathrm{ddir}_{it}
+\beta_{\text{front}}\,\mathrm{FCRP}_{t}\,I^{\text{dec}}_{i}
+\beta_{\text{rear}}\,\mathrm{RCRP}_{t}\,I^{\text{acc}}_{i}
\\
&\;
+\beta_{\text{dist}}\frac{1}{D_t}\left(I^{\text{dec}}_{i}+I^{\text{acc}}_{i}\right).
\end{aligned}
\label{eq:spec_linear_8p}
\end{equation}

Deceleration indicators apply to alternatives $\{1,2,3\}$, acceleration indicators to $\{7,8,9\}$, turning indicators to $\{1,3,4,6,7,9\}$, and the inverse distance term enters only for $\{1,2,3,7,8,9\}$.
Collision angle was defined as the azimuth between the pedestrian velocity vector and the line-of-sight vector from the pedestrian to the AV. 
The angle was mapped to front and rear sector directional intensity measures (0 to 1) using the cosine of the angle within each sector. Larger values indicate stronger alignment. 
These directional intensities were then combined with a traffic conflict indicator known as the closing time-to-collision (CTTC) \cite{alhaideri2025cttc}. This is performed to form a collision risk proximity measure that increased when the AV was both directionally salient and close in time. The frontal collision risk proximity (FCRP) and rear collision risk proximity (RCRP)ariables are computed as
\begin{equation}
\mathrm{FCRP}_{t}=\frac{\tau^{\text{front}}_{t}}{1+\mathrm{CTTC}_{t}},
\qquad
\mathrm{RCRP}_{t}=\frac{\tau^{\text{rear}}_{t}}{1+\mathrm{CTTC}_{t}}.
\label{eq:front_rear_crp}
\end{equation}
where $\tau^{\text{front}}_{t}$ is the front sector directional intensity and $\tau^{\text{rear}}_{t}$ is the rear sector directional intensity.
The front sector risk enters only for alternatives $\{1,2,3\}$ because these correspond to deceleration alternatives. 
This allowes the model to capture pedestrian slowing down when the AV approaches from the front. 
The rear sector risk enters only for alternatives $\{7,8,9\}$ because these correspond to acceleration alternatives. This allows the model to capture pedestrian speeding up when the AV approaches from behind.

The attraction-to-destination components $\mathrm{ddist}_{it}$ and $\mathrm{ddir}_{it}$ are defined for each alternative and enter with generic coefficients for all nine alternatives \cite{alhaideri2025cyclistAAP}. 
$\mathrm{ddist}_{it}$ is the distance in meters between the centroid of alternative $i$ and the destination.
The destination is approximated as the pedestrian's last observed position in the trajectory segment. 
$\mathrm{ddir}_{it}$ is the angular deviation in degrees between two lines at the pedestrian's current position at decision step $t$. 
The first line connects the current position to the destination. The second line connects the current position to the centroid of movement alternative $i$.

For the MNL and spatial GEV models (SCL, GSCL, SCNL, and GSCNL), the same systematic utility in Equation~\eqref{eq:spec_linear_8p} is used, but the attraction-to-destination $\mathrm{ddist}_{it}$ and $\mathrm{ddir}_{it}$ are excluded.
These models are estimated in GAUSS~26 via maximum likelihood. Correlation across alternatives is introduced through the allocation parameters $\alpha_{i,im}$ and nesting coefficients $\lambda$ or $\lambda_k$, as summarized in Table~\ref{tab:model_variable_matrix}. In particular, SCL uses contiguity-based allocation (Equation~\eqref{eq:scl_alpha}) with a single nesting coefficient, GSCL replaces contiguity with distance based allocation (Equation~\eqref{eq:gscl_alpha}), and SCNL allows multiple nesting coefficients across nest categories. GSCNL further estimates the allocation pattern from data through the pseudo logit allocation rule (Equation~\eqref{eq:gscnl_alpha}).

ResLogit is estimated in Python using stochastic gradient based optimization. The systematic utility vector $\mathbf{V}_t=(V_{1t},\ldots,V_{Jt})^\top$ from Equation~\eqref{eq:spec_linear_8p} is passed through the residual correction layers in Equation~\eqref{eq:reslogit_layers} to obtain corrected utilities $\mathbf{U}_t$ and probabilities follow Equation~\eqref{eq:reslogit_prob}. 
For estimation, parameters are learned by minimizing the cross entropy loss (equivalently maximizing the multinomial log-likelihood) using the Adam optimizer with gradient clipping (norm $5.0$) and $\ell_2$ weight decay. 
Hyperparameters are tuned using Bayesian optimization (Gaussian process minimization) to target the validation log-likelihood over learning rate, number of epochs, number of residual layers $M$, and weight decay. The selected configuration for the original data model uses a learning rate of $2.54\times 10^{-2}$, $\ell_2$ weight decay of $10^{-2}$, $M=2$ residual layers, and $500$ training epochs.

\section{Results and Discussions}

Table~\ref{tab:model_fit_all} presents the fit metrics for the ResLogit and spatial GEV models (MNL, SCL, GSCL, SCNL, GSCNL) on the training sample ($N=1{,}850$).
As can be seen from the table, the improvement over the MNL are modest. The mean log-likelihood (LL) increases from $-2.14696$ (MNL) to $-2.13728$ (GSCL), and AIC decreases from $7{,}955.75$ to $7{,}923.94$.
This implies that adding analyst specified proximity correlation ($\alpha$ and $\lambda$ structures) may provide limited additional explanatory power for this dense $3\times3$ movement grid under the current sample size.
The ResLogit achieves substantially higher fit, with mean log-likelihood $-1.71634$ and AIC of $6{,}690.47$.

This improvement in fit reflects the inclusion of the learned residual cross effects that adjust utilities beyond the linear component.
The magnitude of the fit differences is also informative. The spatial GEV models improves mean log likelihood by less than $0.01$ relative to MNL.
The ResLogit improves fit by roughly $0.43$ per observation. 
In dense discrete movement grids, small mean log-likelihood changes can translate into noticeable probability re-allocation between neighbouring alternatives. 
The confusion matrices in Figure~\ref{fig:gev_cm_maxp} indicate that the GEV structures did not meaningfully change predicted substitution patterns in this setting. 
Hence, the ResLogit improvement is not only a statistical artefact. It corresponded to a qualitative change in predictive behaviour where probability allocation becomes less concentrated in a few dominant classes and more consistently distributed across neighbouring alternatives (Figure~\ref{fig:cm_all}).

\begin{table}[H]
\centering
\caption{Model fit on the estimation sample (original train set, $N=1{,}850$).}
\label{tab:model_fit_all}
\small
\setlength{\tabcolsep}{5pt}
\begin{tabular}{l r r r r }
\toprule
\multicolumn{1}{l}{Model} &
\multicolumn{1}{l}{$k$} &
\multicolumn{1}{l}{LL} &
\multicolumn{1}{l}{Mean LL} &
\multicolumn{1}{l}{AIC} \\

\midrule
MNL   & 6   & -3{,}971.8760 & -2.14696 & 7{,}955.75  \\
SCL   & 7   & -3{,}957.7975 & -2.13935 & 7{,}929.60  \\
GSCL  & 8   & -3{,}953.9680 & -2.13728 & 7{,}923.94 \\
SCNL  & 8   & -3{,}957.7050 & -2.13930 & 7{,}931.41  \\
GSCNL & 9   & -3{,}955.4295 & -2.13807 & 7{,}928.86  \\
\midrule
ResLogit & 8 & -3{,}175.2334 & -1.71634 & 6{,}366.46  \\
\bottomrule
\end{tabular}
\end{table}

A plausible explanation is that the $3\times3$ grid is both dense and highly symmetric. This compresses behavioural variation into small differences between neighbouring alternatives. 
Hence, in such settings, analyst specified correlation structures can become weakly identified. 
The pairwise nest allocation rules in SCL and GSCL impose a particular geometry of dependence. 
With the available sample size and class imbalance, the data may not have been informative enough to clearly favour one spatial linkage design over another.
In addition, overlapping nests distributed correlation across many pairwise components. This can dilute the effective influence of the nesting coefficient on predicted probabilities when the systematic utilities already dominated the ranking of alternatives.
The marginal improvement of the spatial GEV models therefore does not imply that proximity based correlation is absent. 
Rather, they suggest that the specific closed form GEV parameterizations evaluated here, based on predefined contiguity or distance decay linkage rules, are not flexible enough to reproduce the empirical pattern of local substitution in the movement grid. 
In contrast, ResLogit improved fit by learning cross alternative corrections directly on the utility vector. The model learns which neighbouring alternatives are most frequently misclassified relative to the realized movement.

\subsection{Confusion matrices and classification metrics}

\begin{figure*}[!t]
\centering
\newcommand{\cmh}{0.27\textheight}

\begin{tabular}{cc}
\begin{minipage}{0.48\textwidth}\centering
\textbf{(a) SCL}\\[4pt]
\includegraphics[width=\linewidth,height=\cmh,keepaspectratio]{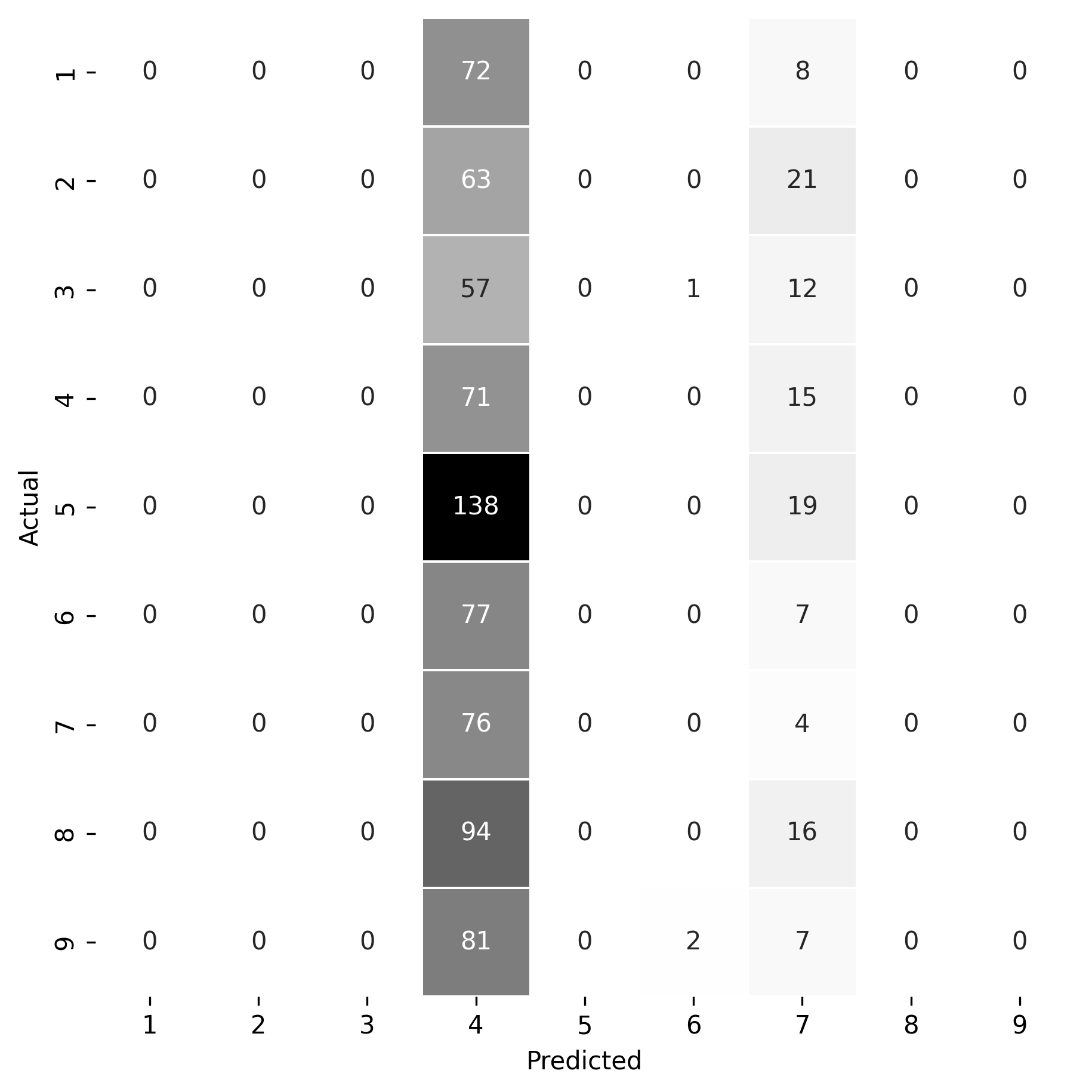}
\end{minipage}
&
\begin{minipage}{0.48\textwidth}\centering
\textbf{(b) GSCL}\\[4pt]
\includegraphics[width=\linewidth,height=\cmh,keepaspectratio]{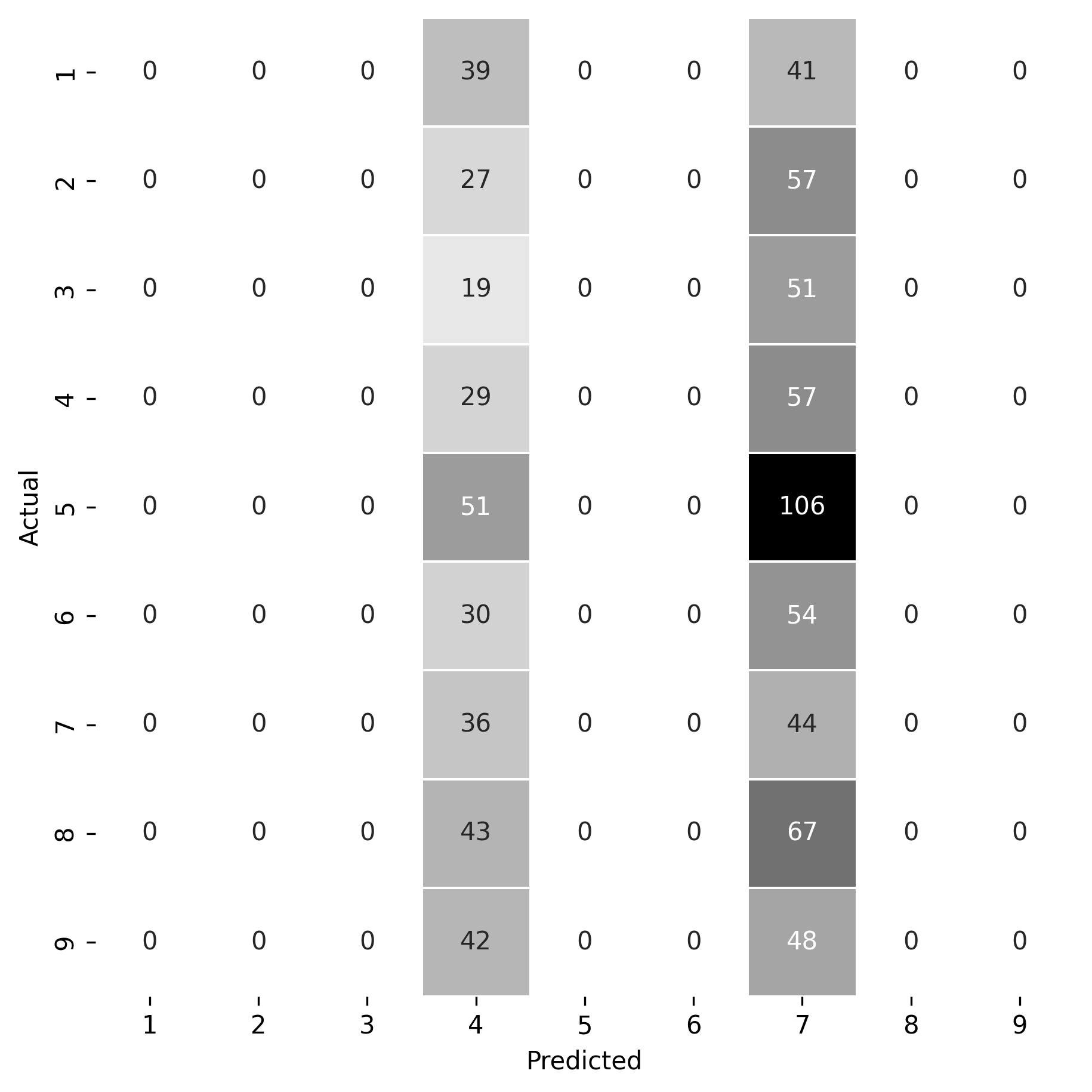}
\end{minipage}
\end{tabular}

\vspace{6pt}

\begin{tabular}{cc}
\begin{minipage}{0.48\textwidth}\centering
\textbf{(c) SCNL}\\[4pt]
\includegraphics[width=\linewidth,height=\cmh,keepaspectratio]{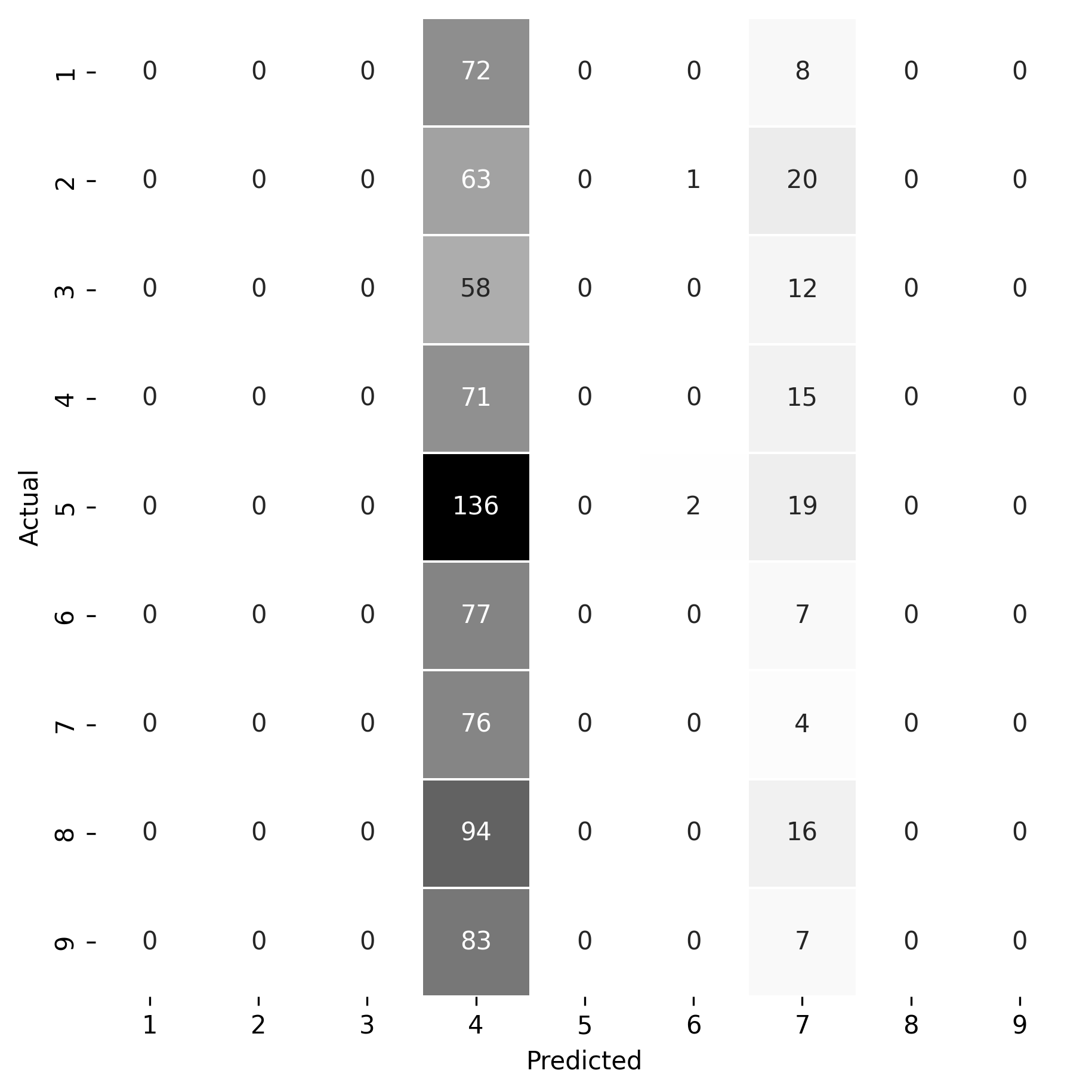}
\end{minipage}
&
\begin{minipage}{0.48\textwidth}\centering
\textbf{(d) GSCNL}\\[4pt]
\includegraphics[width=\linewidth,height=\cmh,keepaspectratio]{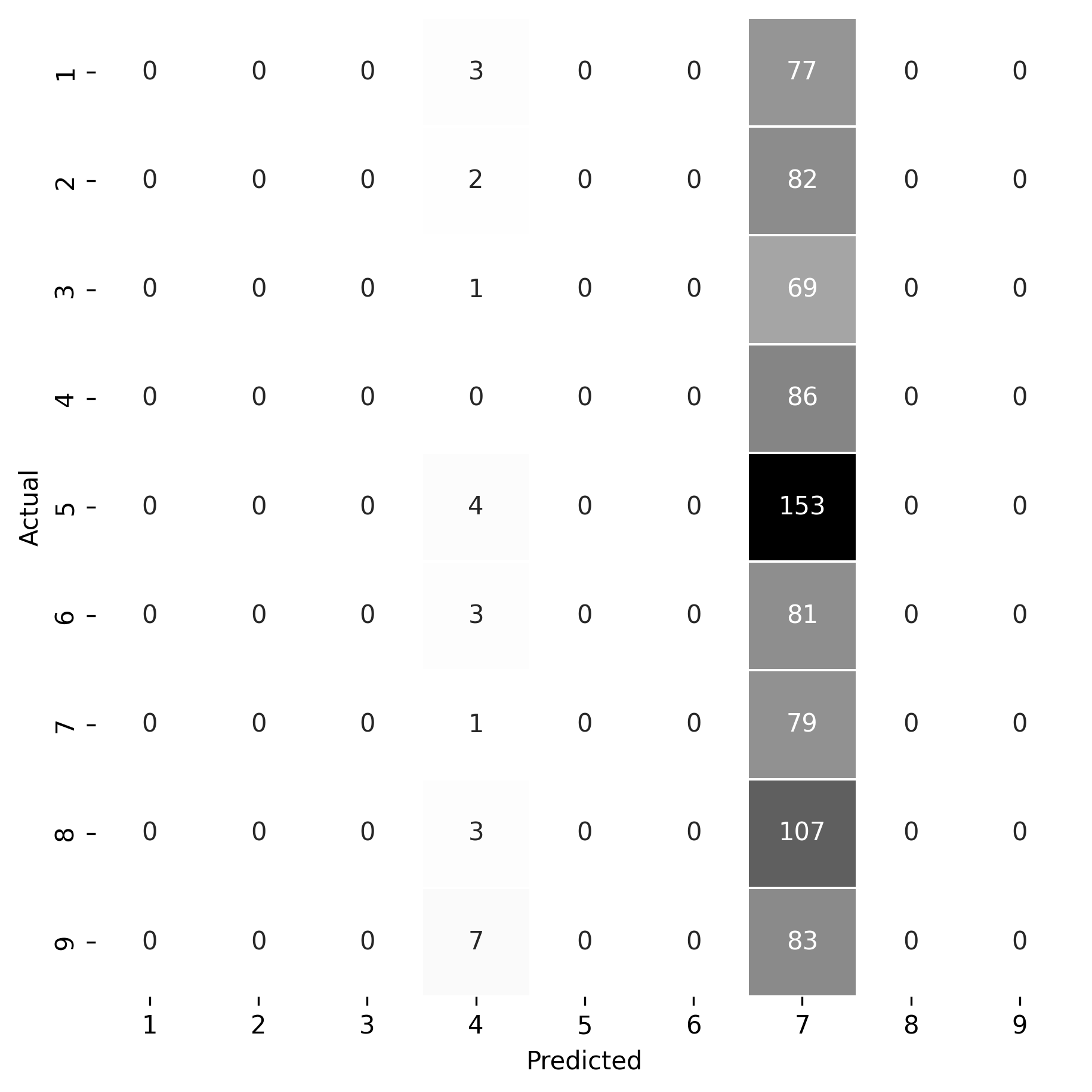}
\end{minipage}
\end{tabular}

\vspace{6pt}

\begin{minipage}{0.48\textwidth}\centering
\textbf{(e) MNL}\\[4pt]
\includegraphics[width=\linewidth,height=\cmh,keepaspectratio]{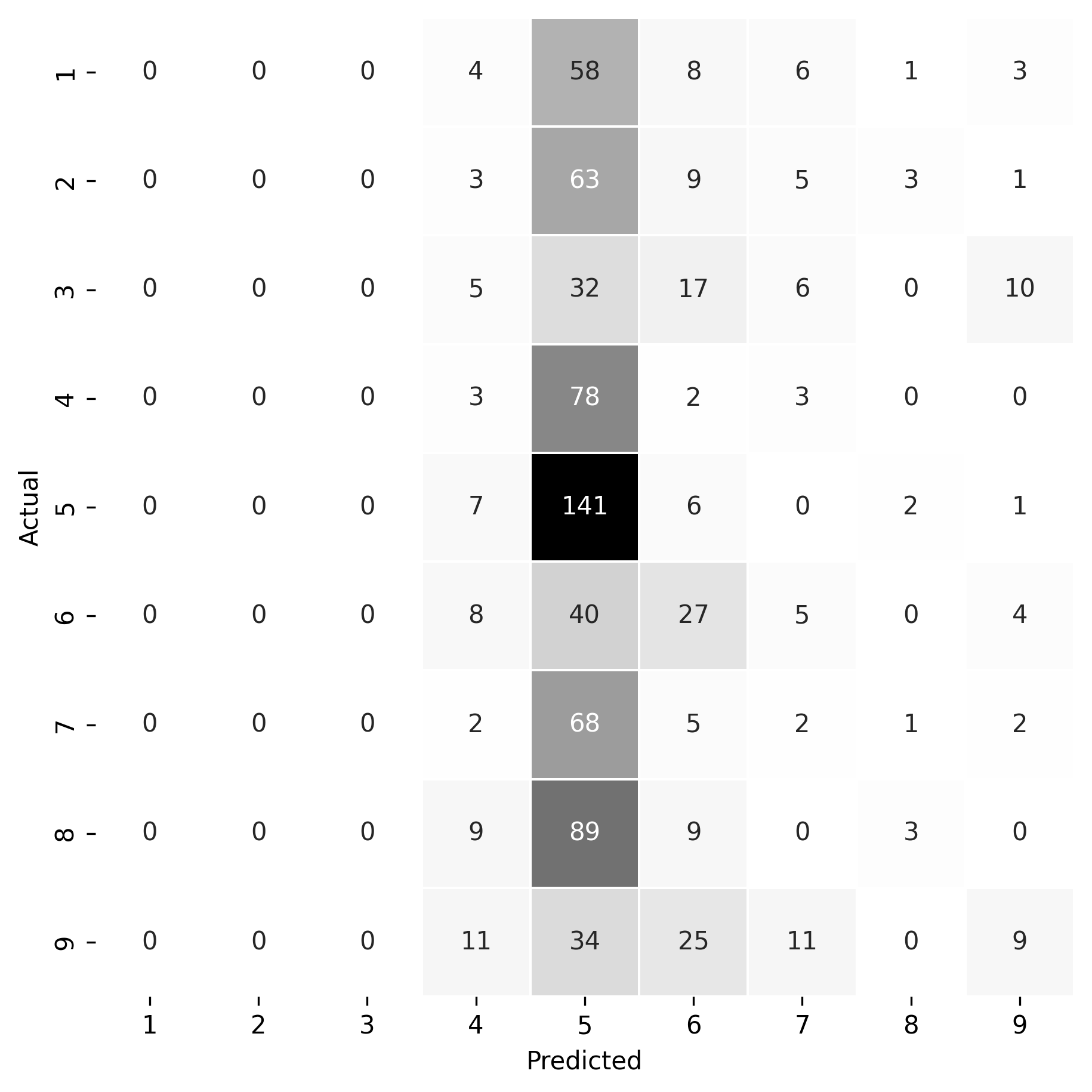}
\end{minipage}

\caption{Confusion matrices for the spatial GEV models using maximum probability (MaxP) predictions.}
\label{fig:gev_cm_maxp}
\end{figure*}

To examine the predictive performance for the spatial GEV models, we obtain the maximum probability (MaxP) predictions. This is calculated by selecting the alternative with the largest predicted choice probability at each decision step.
Figure~\ref{fig:gev_cm_maxp} shows that the GEV specifications concentrate predictions in a small subset of alternatives which collapses much of the $3\times3$ grid into dominant choices.
This is consistent with the marginal improvements in Table~\ref{tab:model_fit_all}. The estimated correlation structure is not strong enough (or not identifiable enough) to materially change predicted substitution patterns in this setting relative to the MNL baseline.

In contrast, the ResLogit model produces more distributed predictions (Figure~\ref{fig:cm_all}), with most errors occurring among neighbouring movement alternatives. This error structure is behaviourally consistent with local similarity in the movement grid.
Table~\ref{tab:reslogit_metrics} quantifies this pattern. The top-1 accuracy is $0.3557$ on the original training set and $0.3215$ on the original test set. Also, the top-3 accuracy increases up to $0.7438$ (train) and $0.6714$ (test). The balanced accuracy and macro-F1 remain lower than weighted metrics which reflects class imbalance and uneven recall across the nine alternatives.

From an AV planning perspective, the structure of prediction errors matters. 
Mis-classifying the exact cell in a discretized grid is not equally harmful among all errors. 
Confusions concentrated among adjacent cells correspond to small deviations in speed adjustment or heading change. This could be more consistent with human movement variability and sensing noise. 
In contrast, errors that jump across the grid correspond to qualitatively different actions. 
The ResLogit confusion matrices show that most errors remained local. This is consistent with the hypothesis that the residual layers learned a proximity consistent correction to the MNL.

Several additional observations are also noted from Table~\ref{tab:reslogit_metrics}. 
First, the moderate top-1 accuracy indicated that the exact choice prediction remained difficult under naturalistic variability. This is expected when neighbouring alternatives differ only by small micro adjustments. 
Second, the substantial gap between top-1 and top-3 accuracy is informative. 
Although the exact cell is not always predicted, the model frequently assigned high probability to a small local set of plausible neighbouring movements. 
This behaviour could be operationally meaningful for AV systems that evaluate multiple candidate pedestrian motions rather than relying on a single point prediction. 

Third, the lower balanced accuracy and macro-F1 indicated that minority movement classes are more difficult to recover which is consistent with limited representation of rare alternatives.
Finally, the synthetic results served as an internal consistency check. 
When features are generated with increased variability and choices are sampled from the fitted model, the performance increases sharply (top-1 above $0.80$). 
This indicats that the ResLogit specification could recover strong decision patterns when sufficient heterogeneity is present in the data. 
The lower performance on the naturalistic sample could more plausibly reflect limited behavioural separation between neighbouring alternatives and class imbalance, rather than numerical instability or model mis-specification.

\begin{table}[t]
\centering
\caption{ResLogit predictive metrics on original and synthetic datasets (train and test).}
\label{tab:reslogit_metrics}
\small
\setlength{\tabcolsep}{3pt}
\begin{adjustbox}{max width=\linewidth}
\begin{tabular}{ll r r r r r r r r}
\toprule
Dataset & Split & $N$ & Mean LL & Top-1 & Top-2 & Top-3 & Bal. acc & F1 (macro) & F1 (wtd) \\
\midrule
Original  & Train & 1{,}850 & -1.71634 & 0.355676 & 0.599459 & 0.743784 & 0.317683 & 0.316760 & 0.329193 \\
Original  & Test  &   843   & -1.84081 & 0.321471 & 0.533808 & 0.671412 & 0.287930 & 0.281386 & 0.296082 \\
\midrule
Synthetic & Train & 35{,}000 & -0.49340 & 0.808971 & 0.950057 & 0.984514 & 0.761628 & 0.767302 & 0.808282 \\
Synthetic & Test  & 15{,}000 & -0.50019 & 0.807467 & 0.948333 & 0.982933 & 0.754620 & 0.762970 & 0.806300 \\
\bottomrule
\end{tabular}
\end{adjustbox}
\end{table}

\subsection{Interpretation of linear utility coefficients}

Table~\ref{tab:reslogit_coef_both} presents the estimated linear utility parameters in Equation~\eqref{eq:spec_linear_8p} for both the processed naturalistic sample (\textit{Original}) and the synthetically generated sample (\textit{Synthetic}). 

\begin{table*}[t]
\centering
\caption{ResLogit linear-utility coefficient estimates.}
\label{tab:reslogit_coef_both}
\small
\setlength{\tabcolsep}{6pt}
\begin{tabular}{lrrrrrr}
\toprule
& \multicolumn{3}{c}{Original (cluster bootstrap)} & \multicolumn{3}{c}{Synthetic (IID bootstrap)} \\
\cmidrule(lr){2-4}\cmidrule(lr){5-7}
Variable & Coef. & S.E. & $p$ & Coef. & S.E. & $p$ \\
\midrule
$\beta_{\text{dec}}$  &  1.091490 & 1.257677 & 0.386576 &  1.010338 & 0.028384 & 0.000000 \\
$\beta_{\text{acc}}$  &  0.471755 & 0.862350 & 0.584988 &  0.453966 & 0.036242 & 0.000000 \\
$\beta_{\text{turn}}$ &  0.266480 & 0.229563 & 0.247188 &  0.226830 & 0.016002 & 0.000000 \\
$\beta_{\text{dist}}$   &  0.033585 & 0.206108 & 0.870735 &  0.018264 & 0.008713 & 0.037334 \\
$\beta_{\text{front}}$ &  0.035592 & 0.327723 & 0.913633 &  0.038941 & 0.012480 & 0.002076 \\
$\beta_{\text{rear}}$  & -0.119108 & 0.449637 & 0.791379 & -0.119021 & 0.014455 & 0.000000 \\
$\beta_{\text{ddist}}$  & -3.177150 & 0.802500 & 0.000107 & -2.996716 & 0.029279 & 0.000000 \\
$\beta_{\text{ddir}}$   & -1.280984 & 0.385602 & 0.001074 & -1.200089 & 0.012576 & 0.000000 \\
\bottomrule
\end{tabular}
\end{table*}

The signs are identical in the two estimates and the magnitudes are close which supports a consistent behavioural interpretation of the linear component.
Across both datasets, the attraction-to-destination terms dominate the behavioural signal. $\beta_{\text{ddist}}$ is negative in both the Original ($-3.177$) and Synthetic ($-2.997$) estimates. 

This implies that alternatives with smaller $\mathrm{ddist}_{it}$ (closer to the destination) yield higher utility. $\beta_{\text{ddir}}$ is also negative in both the Original ($-1.281$) and Synthetic ($-1.200$) estimates. This indicates that alternatives with smaller $\mathrm{ddir}_{it}$ (better alignment with the destination direction) yield higher utility. These two coefficients indicate that pedestrians prefer goal oriented motion, penalizing detours and angular deviation from the destination direction.

The acceleration, deceleration and turn alternative-specific constants capture the general tendency to accelerate, decelerate or turn and have consistent signs across the two datasets. $\beta_{\text{dec}}$ and $\beta_{\text{acc}}$ are positive in both columns which indicates that deceleration ($I^{\text{dec}}_{i}=1$) and acceleration alternatives ($I^{\text{acc}}_{i}=1$) receive a positive baseline utility shift relative to the maintain-speed alternatives. 

The turning constant $\beta_{\text{turn}}$ is also positive in both columns which implies that turning alternatives ($I^{\text{turn}}_{i}=1$) receive a positive utility shift relative to the straight direction reference alternatives albeit close to zero. This aligns with small heading adjustments being common in short horizon pedestrian motion within a discretized $3\times3$ grid.

The interaction terms behave as specified in Equation~\eqref{eq:spec_linear_8p} by affecting only the movement alternatives they are mapped to. The front collision risk proximity term enters as FCRP$_t\,I^{\text{dec}}_{i}$ and is applied only to the deceleration alternatives $\{1,2,3\}$. 
The rear collision risk proximity term enters as RCRP$_t\,I^{\text{acc}}_{i}$ and is applied only to the acceleration alternatives $\{7,8,9\}$. 
The proximity term $\frac{1}{D_t}\left(I^{\text{dec}}_{i}+I^{\text{acc}}_{i}\right)$ also affects only speed change alternatives (deceleration or acceleration). 
The estimated signs follow these mappings and remain consistent across the Original and Synthetic estimates. $\beta_{\text{dist}}>0$ indicates that closer AV distance increases the utility of changing speed. $\beta_{\text{front}}>0$ indicates that higher FCRP$_t$ increases the utility of deceleration alternatives. $\beta_{\text{rear}}<0$ indicates that higher RCRP$_t$ decreases the utility of acceleration alternatives. 
We hypothesize that these variables capture phase dependent perceived urgency during a crossing. 
When the AV is directionally salient in front and closes in time, front collision risk proximity (FCRP$_t$) increases perceived threat and prompts pedestrians to reduce speed, which is consistent with $\beta_{\text{front}}>0$ shifting utility toward deceleration alternatives through FCRP$_t\,I^{\text{front}}_{i}$. 
In contrast, rear collision risk proximity (RCRP$_t$) can be activated in late crossing phases when the interaction is already resolved (for instance, closer to the curb entry after the AV has moved behind the pedestrian). Under this situation, pedestrians experience low urgency to increase speed and tend to maintain their pace rather than accelerate. This is consistent with $\beta_{\text{rear}}<0$ reducing the utility of acceleration alternatives through RCRP$_t\,I^{\text{rear}}_{i}$. Finally, the positive distance sensitivity $\beta_{\text{dist}}>0$ indicates that closer AV distance increases the relative utility of changing speed (deceleration or acceleration) compared to maintaining speed, capturing a generic proximity induced speed adjustment tendency.

\begin{figure*}[!t]
\centering
\newcommand{\cmh}{0.27\textheight}

\begin{tabular}{cc}
\begin{minipage}{0.48\textwidth}\centering
\textbf{(a) Original data (Train)}\\[4pt]
\includegraphics[width=\linewidth,height=\cmh,keepaspectratio]{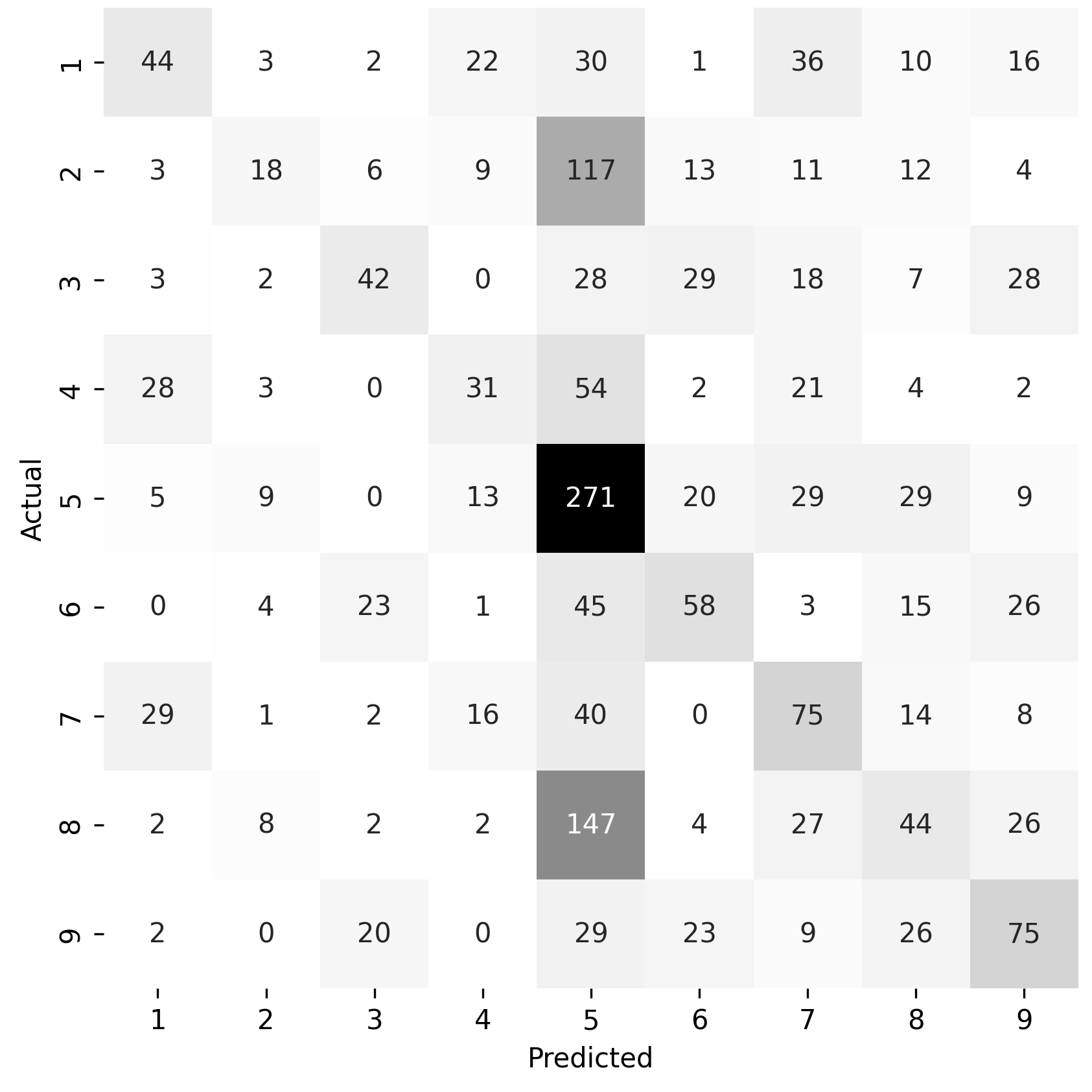}
\end{minipage}
&
\begin{minipage}{0.48\textwidth}\centering
\textbf{(b) Original data (Test)}\\[4pt]
\includegraphics[width=\linewidth,height=\cmh,keepaspectratio]{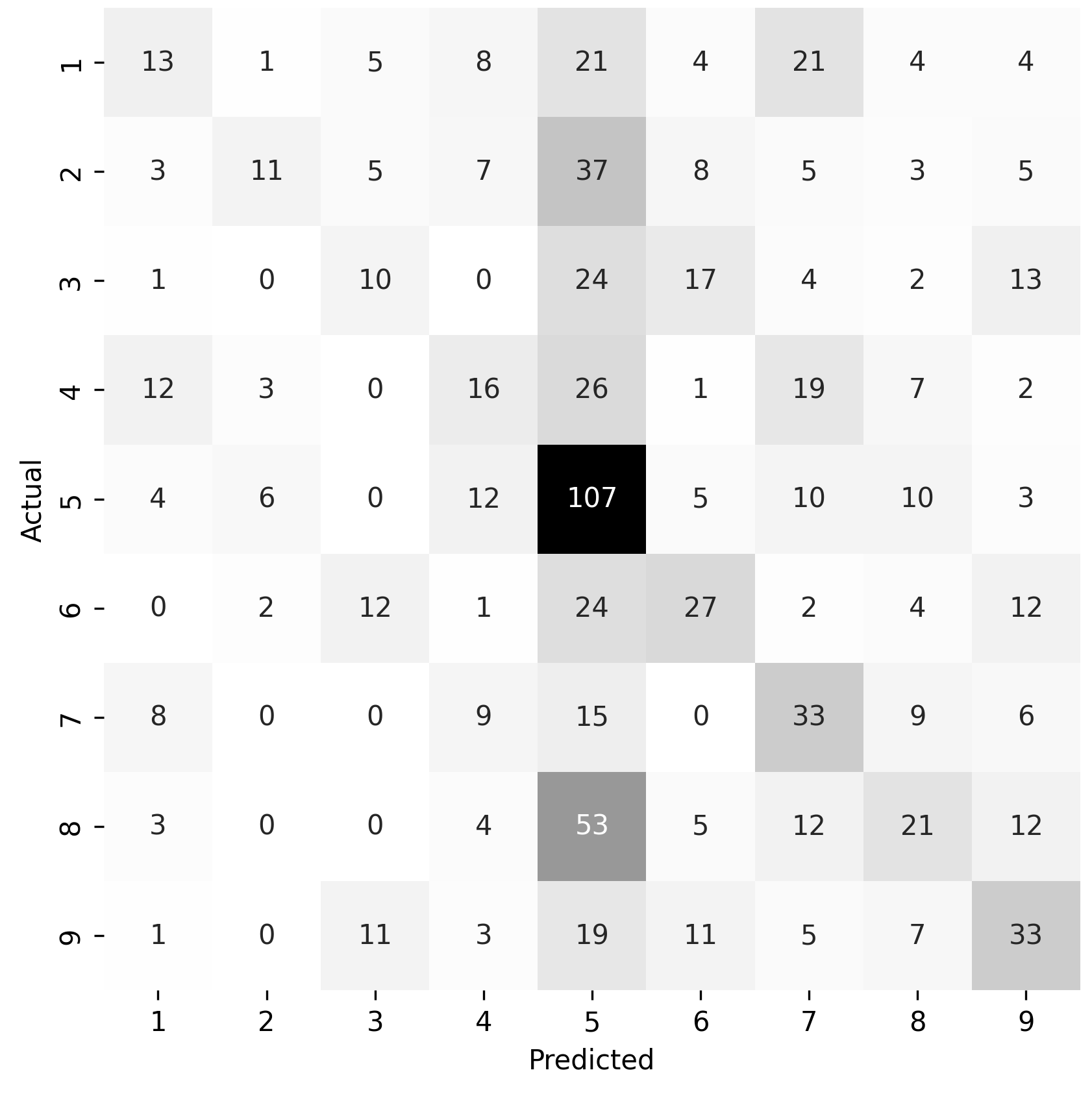}
\end{minipage}
\end{tabular}

\vspace{6pt}

\begin{tabular}{cc}
\begin{minipage}{0.48\textwidth}\centering
\textbf{(c) Synthetic data (Train)}\\[4pt]
\includegraphics[width=\linewidth,height=\cmh,keepaspectratio]{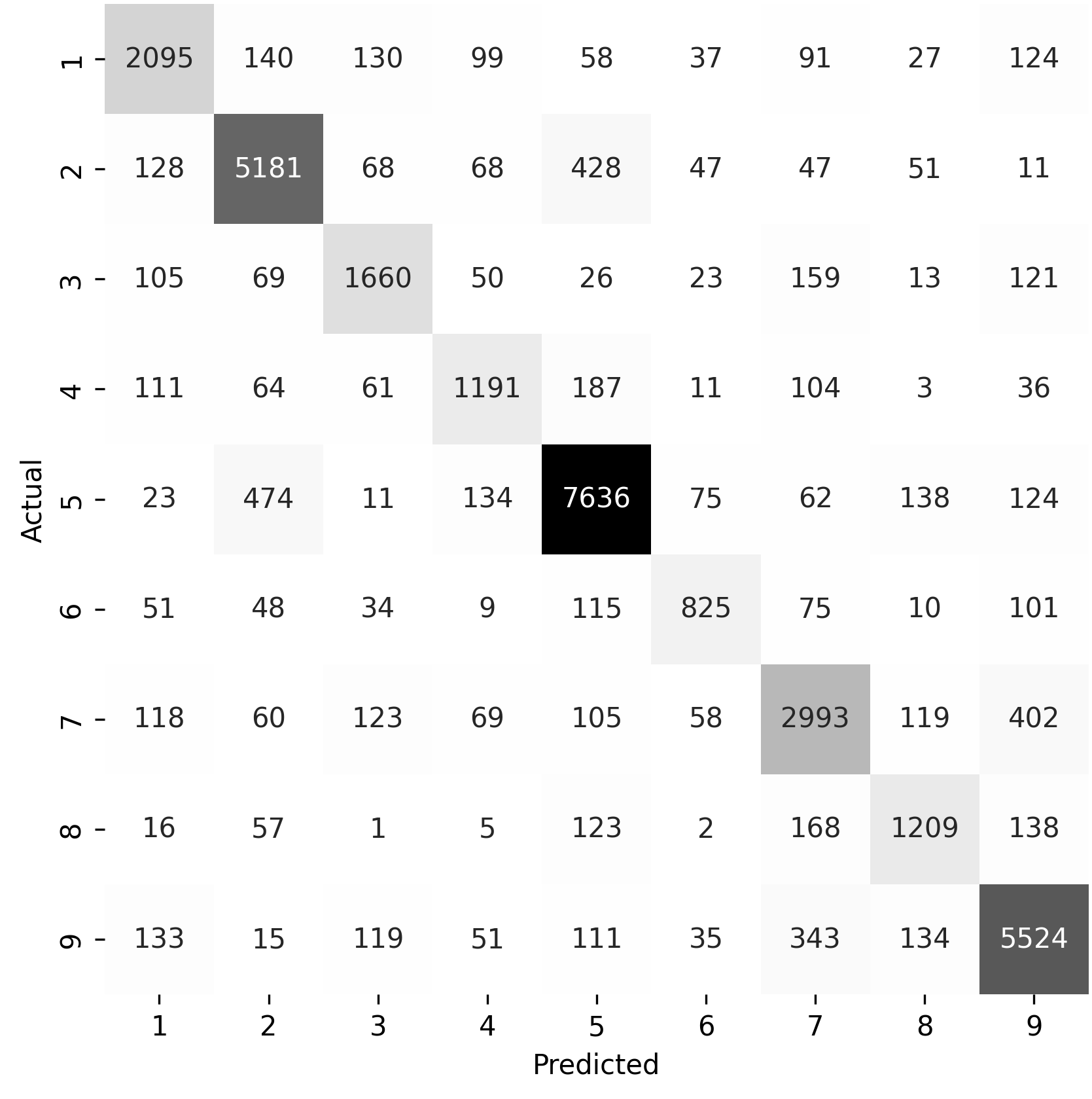}
\end{minipage}
&
\begin{minipage}{0.48\textwidth}\centering
\textbf{(d) Synthetic data (Test)}\\[4pt]
\includegraphics[width=\linewidth,height=\cmh,keepaspectratio]{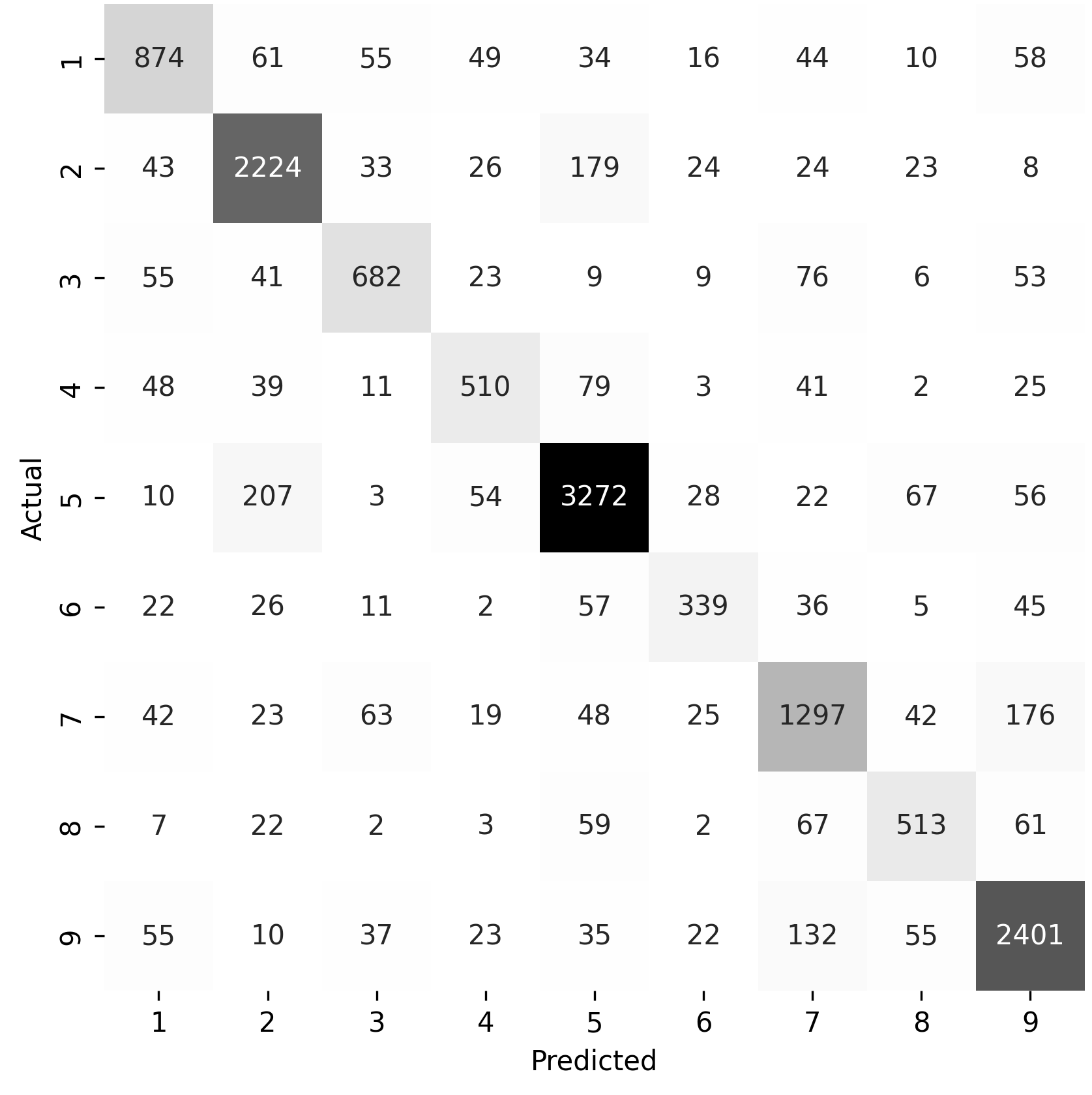}
\end{minipage}
\end{tabular}

\caption{ResLogit confusion matrices on original and synthetic data.}
\label{fig:cm_all}
\end{figure*}

To diagnose whether limited informative variation in the naturalistic sample drives the weaker statistical support observed for several Original data coefficients, we generate a synthetic dataset by sampling features from Normal distributions calibrated to the original training mean and standard deviation, with variance inflated by $\texttt{SIGMA\_MULT}=1.5$. We enforce indicator constraints and clamp continuous variables to admissible ranges. We then sample synthetic choices from the fitted ResLogit probabilities evaluated on these synthetic inputs. This experiment functions as an internal consistency diagnostic and not external validation. Under this synthesis, higher accuracy and stronger statistical significance are expected because variables are generated from the same model class used for re-estimation.
Nevertheless, the synthetic results support one specific conclusion. The weaker significance of several Original data coefficients is plausibly driven by limited heterogeneity in the sample rather than numerical instability or an estimation bug. 
Consistent with this, the synthetic re-estimation yields sharply reduced standard errors and highly significant coefficients, with substantially more diagonal confusion matrices (Figure~\ref{fig:cm_all}) and large increases in balanced accuracy and macro-F1 (Table~\ref{tab:reslogit_metrics}).

\begin{figure*}[!t]
\centering

\begin{tabular}{cc}
\begin{minipage}{0.49\textwidth}\centering
\textbf{(a)} \texttt{inv\_dist}\\[2pt]
\includegraphics[width=\linewidth]{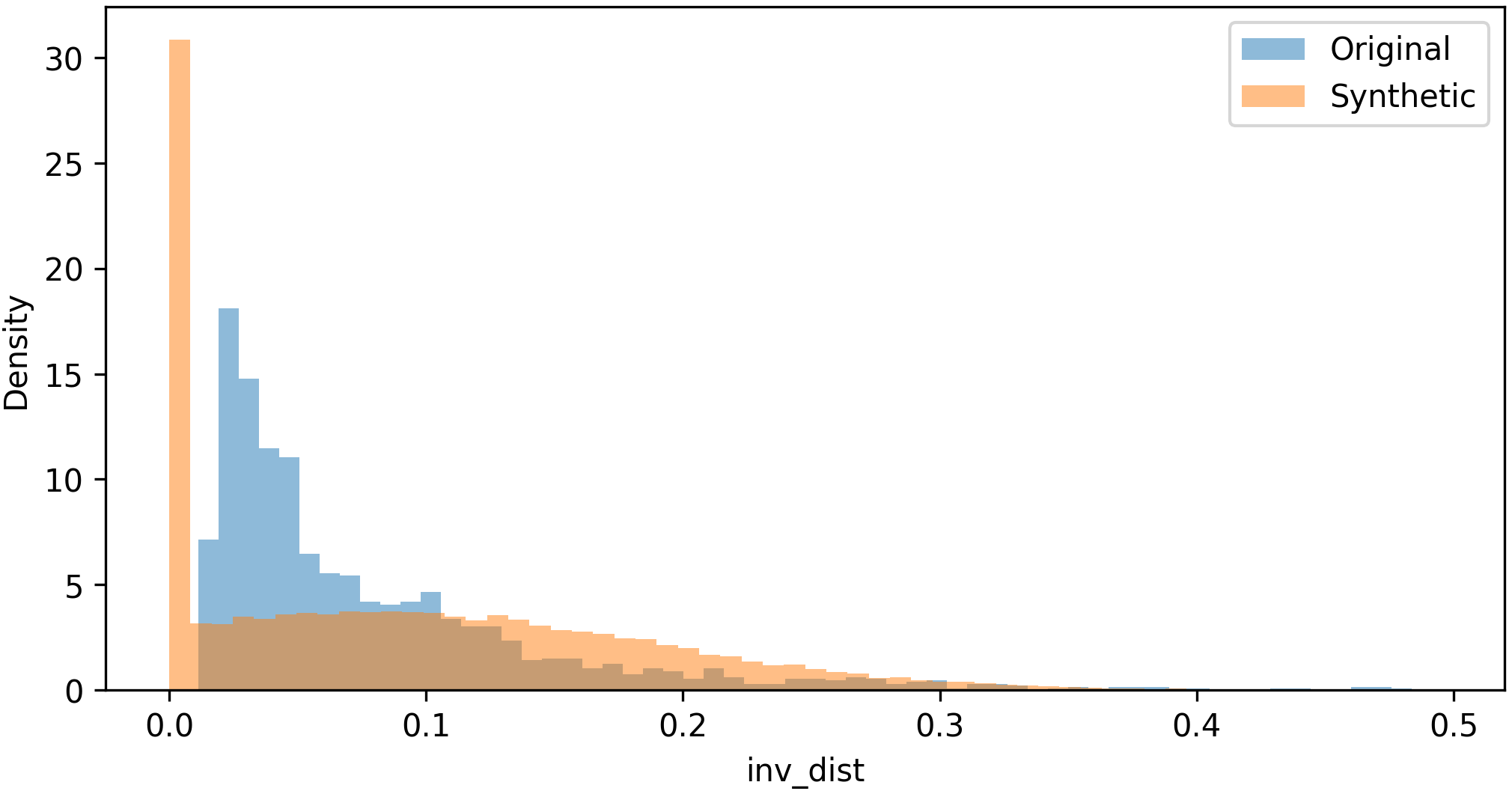}
\end{minipage}
&
\begin{minipage}{0.49\textwidth}\centering
\textbf{(b)} \texttt{FCRP}\\[2pt]
\includegraphics[width=\linewidth]{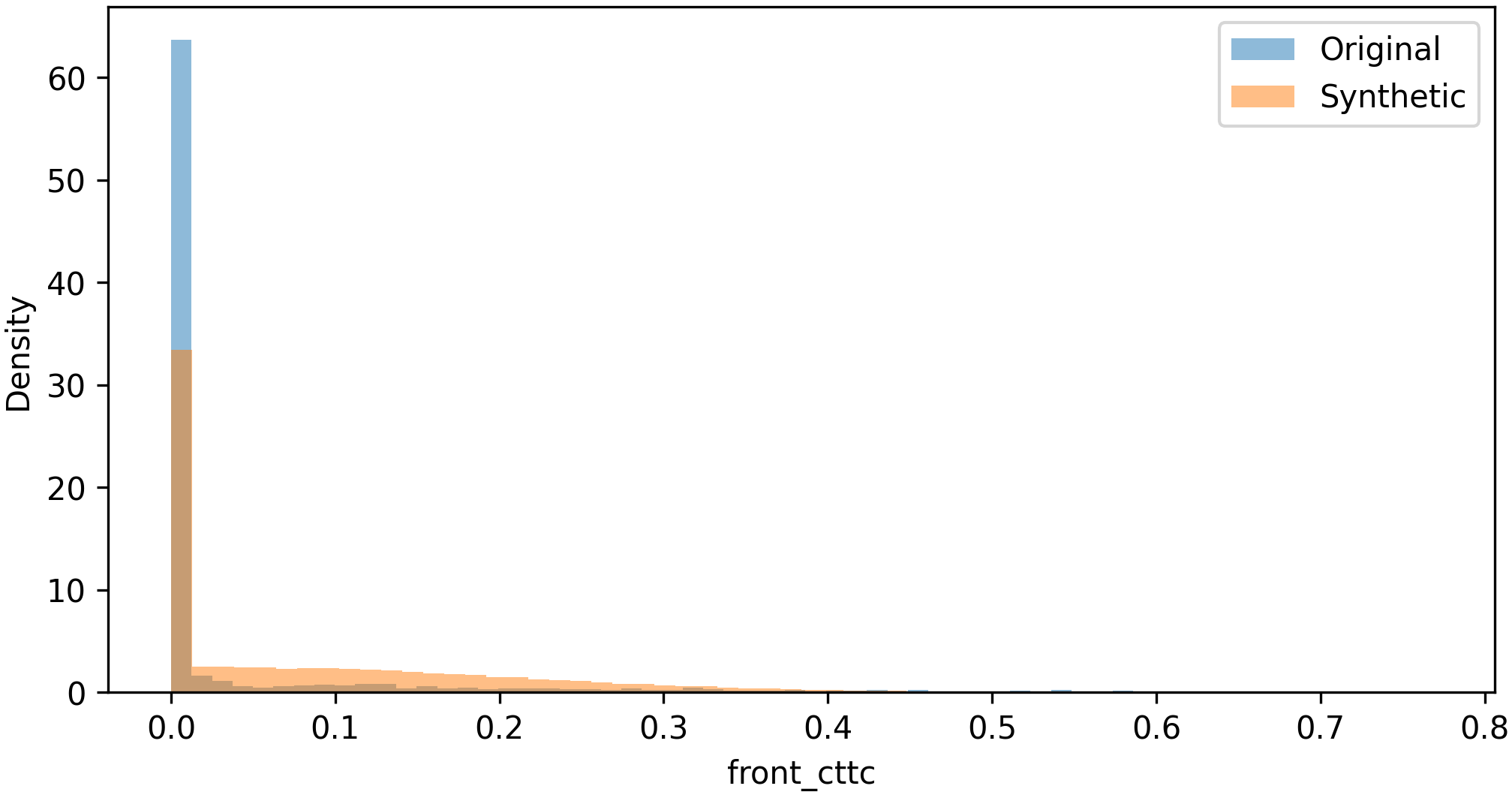}
\end{minipage}
\end{tabular}

\vspace{8pt}

\begin{tabular}{cc}
\begin{minipage}{0.49\textwidth}\centering
\textbf{(c)} \texttt{RCRP}\\[2pt]
\includegraphics[width=\linewidth]{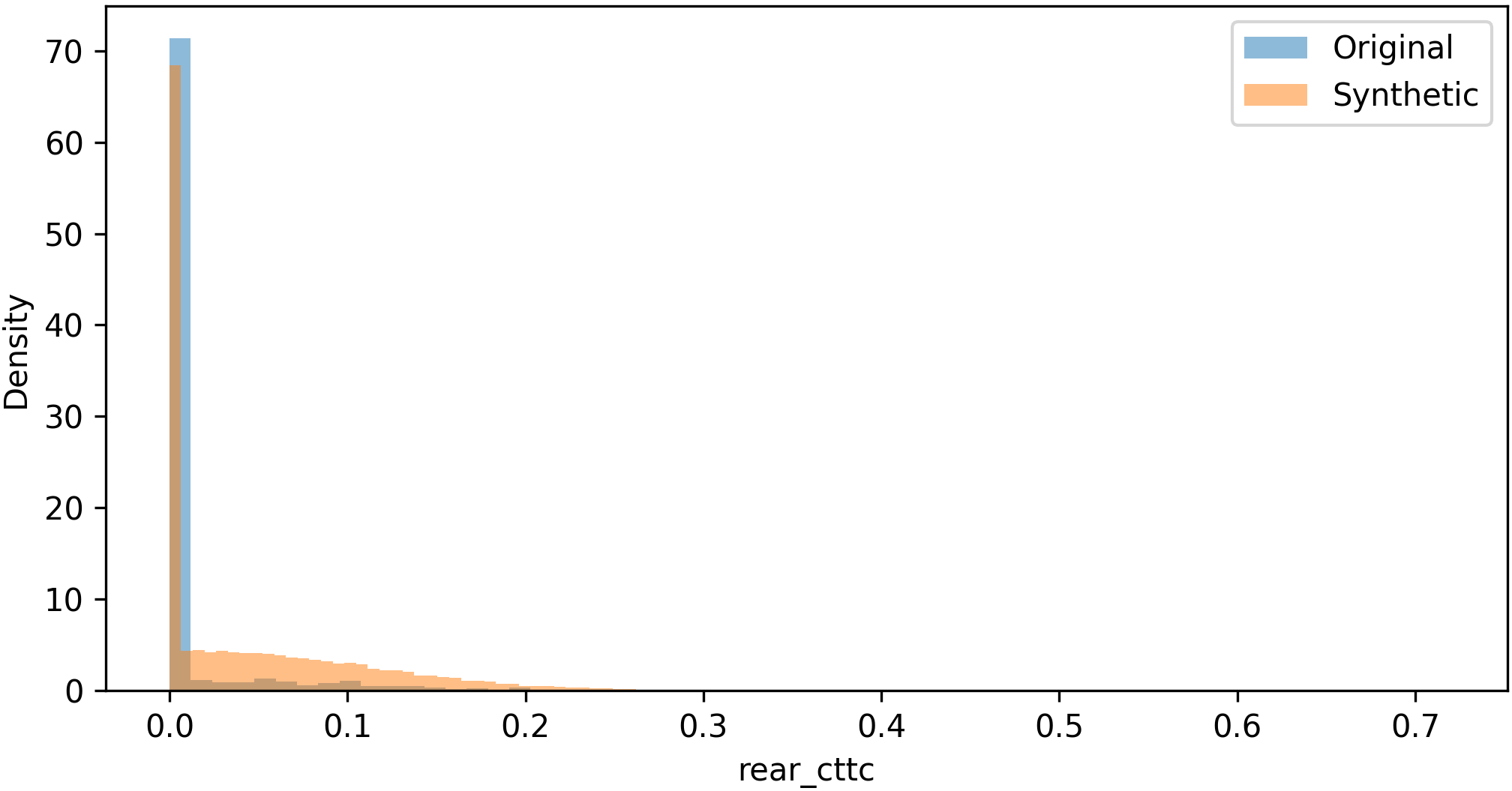}
\end{minipage}
&
\begin{minipage}{0.49\textwidth}\centering
\textbf{(d)} \texttt{ddist} pooled\\[2pt]
\includegraphics[width=\linewidth]{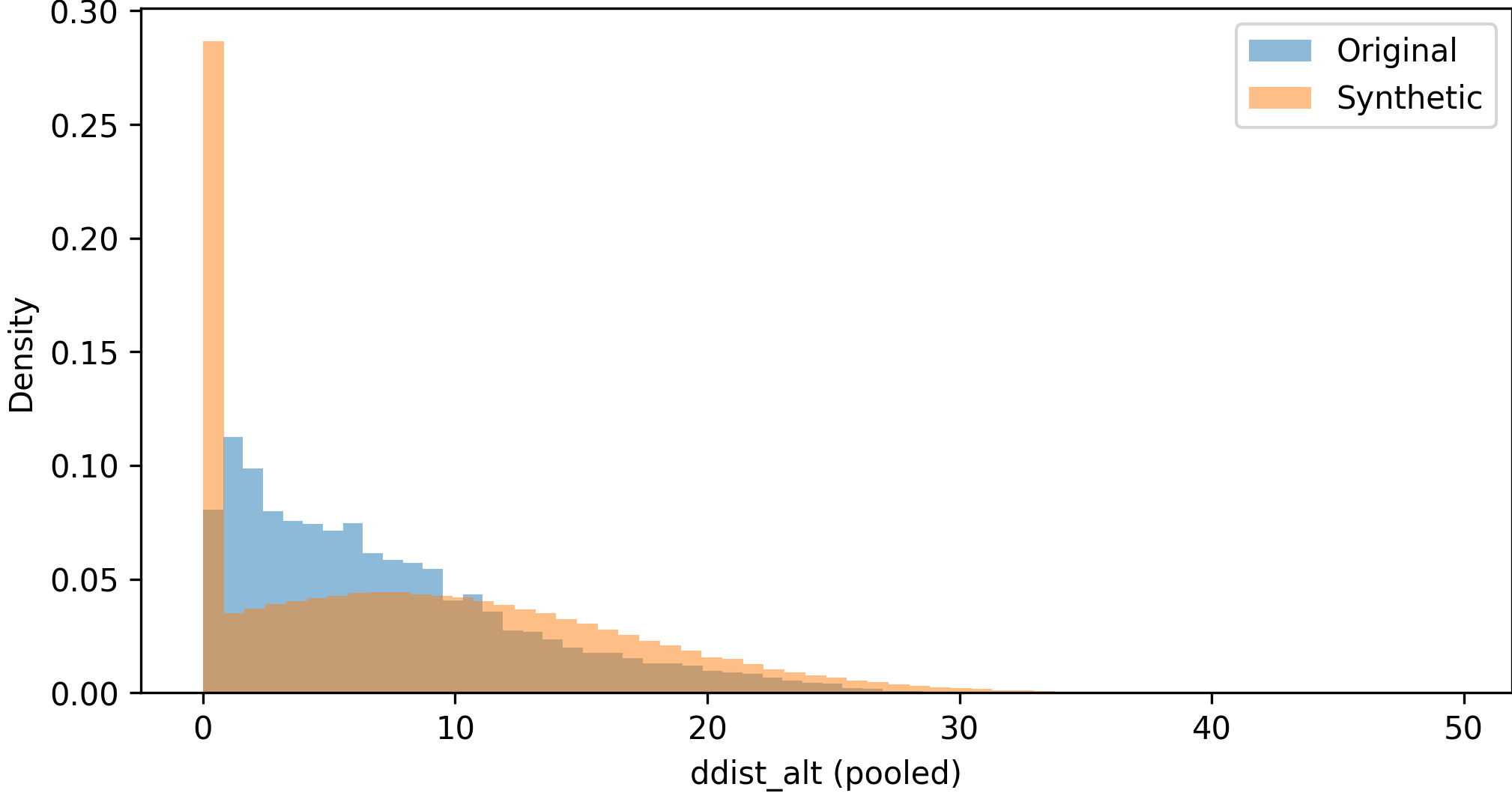}
\end{minipage}
\end{tabular}

\vspace{8pt}

\begin{minipage}{0.60\textwidth}\centering
\textbf{(e)} \texttt{ddir} pooled\\[2pt]
\includegraphics[width=\linewidth]{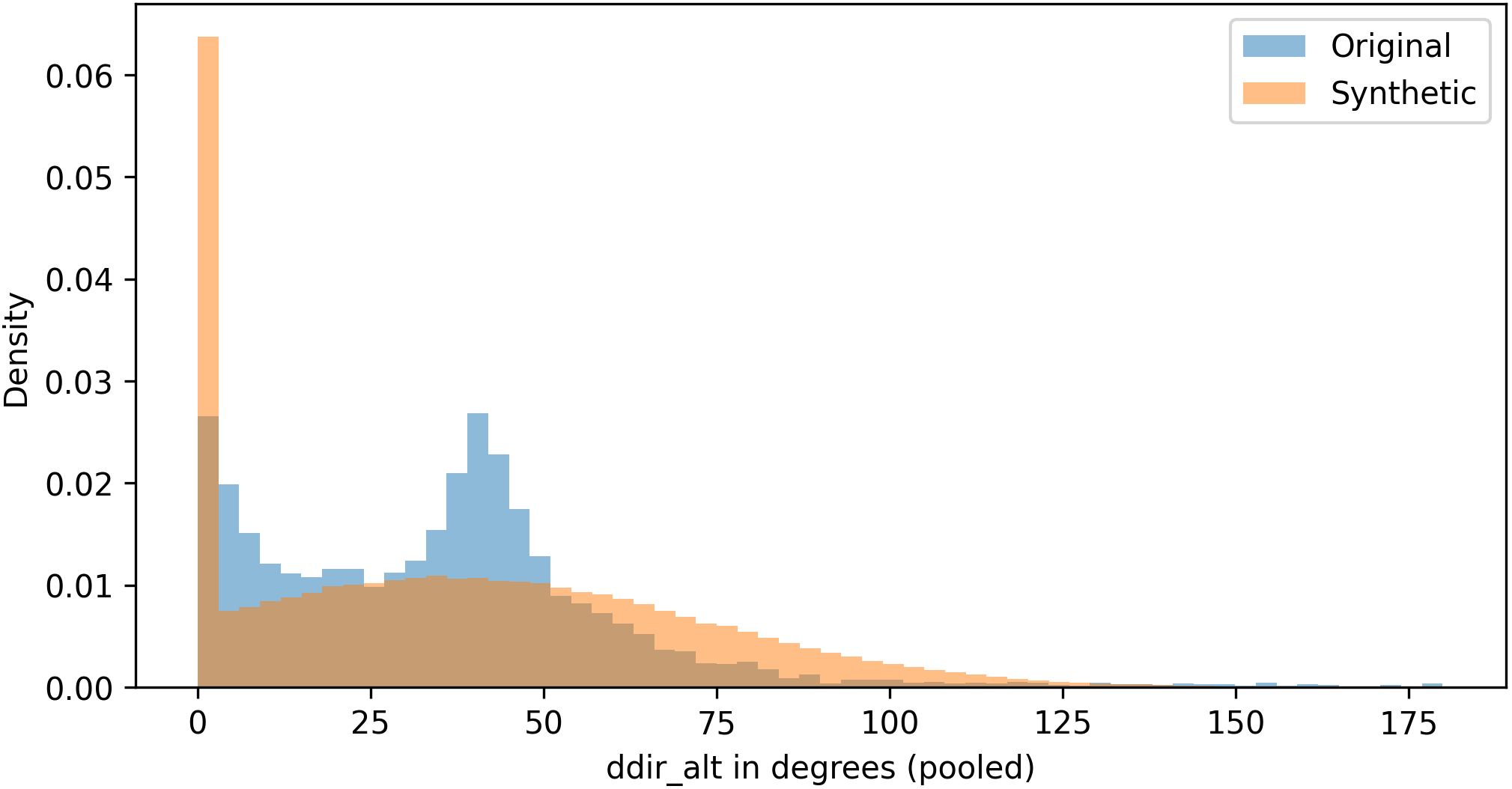}
\end{minipage}

\caption{Original versus synthetic distributions for ResLogit variables. The ddist and ddir plots are pooled across the nine alternatives.}
\label{fig:synth_dists_all}
\end{figure*}

\FloatBarrier

\section{Conclusions}
This paper modelled high frequency next step naturalistic pedestrian motion at midblocks under AV interaction as a spatial discrete choice over a $3\times3$ grid of speed and heading adjustments. Then, a comparison between classic spatial GEV models (SCL, GSCL, SCNL, and GSCNL) against ResLogit is conducted. 
The empirical results show that imposing analyst specified proximity correlation through spatial GEV structures provides only modest gains over an MNL in this dense, high frequency setting (mean log-likelihood improves from $-2.147$ to $-2.137$ at best). 
In contrast, ResLogit substantially improves model fit (mean log-likelihood $-1.716$) and yields error patterns that are concentrated among neighbouring movement alternatives. This is consistent with local similarity in the discretized movement grid. 

Predictive performance remains moderate for exact class (top-1) prediction but improves considerably under top-$k$ evaluation (top-1 $0.321$, top-3 $0.671$ on the test set). This suggests that the model often assigns high probability to a small local set of plausible next step movements even when the exact cell is missed. 
Although ResLogit introduced learnable residual layers, the model remains grounded in a behavioural discrete choice formulation. 
The linear utility retains structurally interpretable parameters and explicit alternative definitions tied to speed and heading adjustments. 
The residual component acted as a correction mechanism that captured cross alternative dependence not well represented by predefined nests. In this sense, this approach differs from trajectory forecasting models that directly regress future coordinates. 
It preserves behavioural interpretability at the level of action trade-offs while leveraging learning to improve fit and local substitution realism.


Based on interpretation of linear parameters of the model, the coefficient estimates suggest a primarily goal directed movement process modulated by interaction pressure. 
The strong negative destination terms indicate that pedestrians tend to maintain moving toward their destination. It reflects the fact that during crossing, pedestrians favour alternatives that reduces both distance to destination and angular deviation. 
The interaction terms shifts the relative utility of speed change alternatives. A closer proximity increases the propensity to adjust speed rather than maintain it. Also, frontal risk increases the attractiveness of deceleration alternatives. 
The rear risk effect is negative for acceleration. This is consistent with reduced urgency to speed up when the vehicle is behind and the interaction is closer to being resolved. 
These interpretations are enabled by the linear utility structure in ResLogit and remain directly expressible in behavioural terms, which is our key motivation for using a discrete choice formulation.

Several limitations constrain the current conclusions. First, the extracted encounters are AV focused and do not explicitly model interactions with other types road users that exist around the pedestrian or richer scene context. This may reduce separability across the nine movement alternatives. 
Second, the framework treats decisions as myopic next step choices and does not represent temporal dependence, memory, or latent intent, despite pedestrian motion being sequential and history dependent. 
Third, the naturalistic sample may suffer from low heterogeneity as demonstrated when the sample was synthesized. This limits stable estimation for minority movement classes and depressing balanced performance metrics. 
Finally, the modelling design captures pedestrian response to the AV but does not represent the two-way coupling in which vehicle motion adapts simultaneously to pedestrian behaviour. 
This study should be interpreted as a baseline. It isolates alternative correlation in a dense movement grid under a consistent utility specification. This allows subsequent extensions to sequential multi-step prediction and multi-agent interaction modelling.
Future work should extend the choice framework to multi-agent settings by incorporating surrounding road users and explicit interaction context. Also, the myopic assumption should be relaxed through temporal state variables or attention/memory mechanisms that capture intent and short horizon dynamics. 
A coupled pedestrian--vehicle modelling approach should be adopted that jointly represents mutual adaptation would further improve behavioural realism and strengthen applicability to planning, yielding, and simulation pipelines for autonomous driving.


\section*{Author contributions}
The authors confirm contribution to the paper as follows: study conception and
design: Al-Haideri, R., Farooq, B.; data curation: Al-Haideri, R., Farooq, B.; analysis and
interpretation of results: Al-Haideri, R., Farooq, B.; draft manuscript
preparation: Al-Haideri, R., Farooq, B.. All authors reviewed the results and approved
the final version of the manuscript.

\section*{Declaration of Conflicting Interests}
The authors declared no potential conflicts of interest with respect to the research, authorship, and/or publication of this article.

\section*{Funding}
The authors disclosed receipt of the following financial support for the research, authorship, and publication of this article: This research was supported by the Canada Research Chair in Disruptive Transportation Technologies and Services (CRC-2022-00480).




\bibliographystyle{unsrtnat}
\bibliography{references}

@article{Tobler1970,
  title   = {A Computer Movie Simulating Urban Growth in the Detroit Region},
  author  = {Tobler, Waldo R.},
  journal = {Economic Geography},
  volume  = {46},
  number  = {2},
  pages   = {234--240},
  year    = {1970},
  doi     = {10.2307/143141}
}

@book{Train2009,
  title     = {Discrete Choice Methods with Simulation},
  author    = {Train, Kenneth E.},
  publisher = {Cambridge University Press},
  address   = {Cambridge},
  year      = {2009}
}

@article{BolducFortinGordon1997,
  title   = {Multinomial Probit Estimation of Spatially Interdependent Choices: An Empirical Comparison of Two New Techniques},
  author  = {Bolduc, Denis and Fortin, Bernard and Gordon, Stephen},
  journal = {International Regional Science Review},
  volume  = {20},
  number  = {1-2},
  pages   = {77--101},
  year    = {1997},
  doi     = {10.1177/016001769702000105}
}

@inproceedings{AlHaideriFarooq2026PedAV,
  title        = {Modelling Pedestrian Behaviour in Autonomous Vehicle Encounters Using Naturalistic Dataset},
  author       = {Al-Haideri, Rulla and Farooq, Bilal},
  year         = {2026},
  booktitle    = {The 37th IEEE Intelligent Vehicles Symposium},
  eprint       = {2602.05142},
  archivePrefix= {arXiv},
  primaryClass = {physics.soc-ph},
  doi          = {10.48550/arXiv.2602.05142}
}

@article{BhatGuo2004,
  title   = {A Mixed Spatially Correlated Logit Model: Formulation and Application to Residential Choice Modeling},
  author  = {Bhat, Chandra R. and Guo, Jessica},
  journal = {Transportation Research Part B: Methodological},
  volume  = {38},
  number  = {2},
  pages   = {147--168},
  year    = {2004},
  doi     = {10.1016/S0191-2615(03)00005-5}
}

@article{MiyamotoEtAl2004,
  title   = {Discrete Choice Model with Structuralized Spatial Effects for Location Analysis},
  author  = {Miyamoto, Kenji and Vichiensan, V. and Shimomura, Naoki and P{\'a}ez, Antonio},
  journal = {Transportation Research Record},
  volume  = {1898},
  number  = {1},
  pages   = {183--190},
  year    = {2004},
  doi     = {10.3141/1898-22}
}

@article{Sener2011,
  title   = {Accommodating Spatial Correlation Across Choice Alternatives in Discrete Choice Models: An Application to Modeling Residential Location Choice Behavior},
  author  = {Sener, Ipek N. and Pendyala, Ram M. and Bhat, Chandra R.},
  journal = {Journal of Transport Geography},
  volume  = {19},
  number  = {2},
  pages   = {294--303},
  year    = {2011},
  doi     = {10.1016/j.jtrangeo.2010.06.015}
}

@article{PerezLopez2022,
  title   = {Spatially Correlated Nested Logit Model for Spatial Location Choice},
  author  = {P{\'e}rez-L{\'o}pez, Jose-Benito and Novales, Margarita and Orro, Alfonso},
  journal = {Transportation Research Part B: Methodological},
  volume  = {161},
  pages   = {1--12},
  year    = {2022},
  doi     = {10.1016/j.trb.2022.05.006}
}

@article{WeissHabib2017,
  title   = {Examining the Difference Between Park and Ride and Kiss and Ride Station Choices Using a Spatially Weighted Error Correlation (SWEC) Discrete Choice Model},
  author  = {Weiss, Adam and Habib, Khandker Nurul},
  journal = {Journal of Transport Geography},
  volume  = {59},
  pages   = {111--119},
  year    = {2017},
  doi     = {10.1016/j.jtrangeo.2017.01.010}
}

@article{WongFarooq2021,
  title   = {ResLogit: A Residual Neural Network Logit Model for Data-Driven Choice Modelling},
  author  = {Wong, Melvin and Farooq, Bilal},
  journal = {Transportation Research Part C: Emerging Technologies},
  volume  = {126},
  pages   = {103050},
  year    = {2021},
  doi     = {10.1016/j.trc.2021.103050}
}

@article{KamalFarooq2024,
  title   = {Ordinal-ResLogit: Interpretable Deep Residual Neural Networks for Ordered Choices},
  author  = {Kamal, Kimia and Farooq, Bilal},
  journal = {Journal of Choice Modelling},
  volume  = {50},
  pages   = {100454},
  year    = {2024},
  doi     = {10.1016/j.jocm.2023.100454}
}

@inproceedings{Alahi2016SocialLSTM,
  title     = {Social {LSTM}: Human Trajectory Prediction in Crowded Spaces},
  author    = {Alahi, Alexandre and Goel, Kratarth and Ramanathan, Vignesh and Robicquet, Alexandre and Fei-Fei, Li and Savarese, Silvio},
  booktitle = {Proceedings of the IEEE Conference on Computer Vision and Pattern Recognition (CVPR)},
  pages     = {961--971},
  year      = {2016}
}

@inproceedings{Gupta2018SocialGAN,
  title     = {Social {GAN}: Socially Acceptable Trajectories with Generative Adversarial Networks},
  author    = {Gupta, Agrim and Johnson, Justin and Fei-Fei, Li and Savarese, Silvio and Alahi, Alexandre},
  booktitle = {Proceedings of the IEEE Conference on Computer Vision and Pattern Recognition (CVPR)},
  pages     = {2255--2264},
  year      = {2018}
}

@inproceedings{Ivanovic2019Trajectron,
  title     = {The Trajectron: Probabilistic Multi-Agent Trajectory Modeling with Dynamic Spatiotemporal Graphs},
  author    = {Ivanovic, Boris and Pavone, Marco},
  booktitle = {Proceedings of the IEEE/CVF International Conference on Computer Vision (ICCV)},
  pages     = {2375--2384},
  year      = {2019}
}

@article{Salzmann2020TrajectronPlusPlus,
  title   = {Trajectron++: Dynamically-Feasible Trajectory Forecasting with Heterogeneous Data},
  author  = {Salzmann, Tim and Ivanovic, Boris and Chakravarty, Prateek and Pavone, Marco},
  journal = {Computer Vision -- ECCV 2020},
  year    = {2020},
  note    = {Lecture Notes in Computer Science (LNCS)}
}

@inproceedings{Gao2020VectorNet,
  title     = {VectorNet: Encoding {HD} Maps and Agent Dynamics from Vectorized Representation},
  author    = {Gao, Jiyang and Sun, Chen and Zhao, Hang and Shen, David and Anguelov, Dragomir and Li, Changqing and Schmid, Cordelia},
  booktitle = {Proceedings of the IEEE/CVF Conference on Computer Vision and Pattern Recognition (CVPR)},
  pages     = {11525--11533},
  year      = {2020}
}

@article{KalatianFarooq2022ContextAware,
  title   = {A context-aware pedestrian trajectory prediction framework for automated vehicles},
  author  = {Kalatian, Arash and Farooq, Bilal},
  journal = {Transportation Research Part C: Emerging Technologies},
  year    = {2022}
}

@article{RasouliTsotsos2019Survey,
  title   = {Autonomous vehicles that interact with pedestrians: A survey of theory and practice},
  author  = {Rasouli, Amir and Tsotsos, John K.},
  journal = {IEEE Transactions on Intelligent Transportation Systems},
  year    = {2019}
}

@article{Landry2024CrossingIntentionReview,
  title   = {Predicting pedestrian crossing intention in autonomous vehicles: A review},
  author  = {Landry, F. G. and others},
  journal = {Neurocomputing},
  year    = {2024}
}

@inproceedings{caesar2020nuscenes,
  title     = {nuScenes: A Multimodal Dataset for Autonomous Driving},
  author    = {Caesar, Holger and Bankiti, Varun and Lang, Alex H. and Vora, Sourabh and Liong, Venice Erin and Xu, Qiang and Krishnan, Anush and Pan, Yu and Baldan, Giancarlo and Beijbom, Oscar},
  booktitle = {Proceedings of the IEEE/CVF Conference on Computer Vision and Pattern Recognition (CVPR)},
  pages     = {11621--11631},
  year      = {2020}
}

@inproceedings{Wilson2021Argoverse2,
  title     = {Argoverse 2: Next Generation Datasets for Self-Driving Perception and Forecasting},
  author    = {Wilson, Benjamin and Qi, William and Agarwal, Tanmay and Lambert, John and Singh, Jagjeet and Khandelwal, Siddhesh and Pan, Bowen and Kumar, Ratnesh and Hartnett, Andrew and Pontes, Jhony Kaesemodel and Ramanan, Deva and Carr, Peter and Hays, James},
  booktitle = {Proceedings of the Neural Information Processing Systems Track on Datasets and Benchmarks},
  year      = {2021}
}

@article{alhaideri2025cttc,
  title   = {Beyond Traditional Models: A Discrete Choice Approach to Vehicle Interaction Modelling at Roundabouts},
  author  = {Al-Haideri, Rulla and Ismail, Karim and Weiss, Adam},
  journal = {Transportmetrica A: Transport Science},
  pages   = {1--45},
  year    = {2025}
}

@article{alhaideri2025cyclistAAP,
  title   = {A Discrete Choice Latent Class Method for Capturing Unobserved Heterogeneity in Cyclist Crossing Behaviour at Crosswalks},
  author  = {Al-Haideri, Rulla and Weiss, Adam and Ismail, Karim},
  journal = {Accident Analysis \& Prevention},
  volume  = {211},
  pages   = {107850},
  year    = {2025}
}

@unpublished{AlHaideriGSCNL,
  title  = {Cyclists' Crossing Behaviour at Roundabouts: A Generalized Spatially Correlated Nested Logit Model},
  author = {Al-Haideri, Rulla and Ismail, Karim and Weiss, Adam},
  note   = {Under review},
  year   = {2026}
}

@article{WongFarooq2021ResLogit,
  title={ResLogit: A residual neural network logit model for data-driven choice modelling},
  author={Wong, Melvin and Farooq, Bilal},
  journal={Transportation Research Part C: Emerging Technologies},
  volume={126},
  pages={103050},
  year={2021},
  publisher={Elsevier}
}

@inproceedings{PhanMinh2020CoverNet,
  title={CoverNet: Multimodal Behavior Prediction Using Trajectory Sets},
  author={Phan-Minh, Tung and Grigore, Elena Corina and Boulton, Freddy A and Beijbom, Oscar and Wolff, Eric M},
  booktitle={Proceedings of the IEEE/CVF Conference on Computer Vision and Pattern Recognition (CVPR)},
  year={2020}
}

@article{Chai2020MultiPath,
  title={MultiPath: Multiple Probabilistic Anchor Trajectory Hypotheses for Behavior Prediction},
  author={Chai, Yuning and Sapp, Benjamin and Bansal, Mayank and Anguelov, Dragomir},
  journal={Conference on Learning Theory (COLT)},
  year={2020}
}

@inproceedings{Biktairov2020PRANK,
  title={PRANK: Motion Prediction via Ranking},
  author={Biktairov, Yevgeny and Zholus, Artem and Kobzev, Alexey and Gotovets, Alexander and Nikolenko, Sergey},
  booktitle={Advances in Neural Information Processing Systems (NeurIPS)},
  year={2020}
}

@article{AntoniniBierlaireWeber2006,
  title = {Discrete choice models of pedestrian walking behavior},
  author = {Antonini, Gianluca and Bierlaire, Michel and Weber, Mats},
  journal = {Transportation Research Part B: Methodological},
  volume = {40},
  number = {8},
  pages = {667--687},
  year = {2006},
  doi = {10.1016/j.trb.2005.09.006}
}

@article{RobinAntoniniBierlaireCruz2009,
  title = {Specification, estimation and validation of a pedestrian walking behavior model},
  author = {Robin, Th. and Antonini, G. and Bierlaire, M. and Cruz, J.},
  journal = {Transportation Research Part B: Methodological},
  volume = {43},
  number = {1},
  pages = {36--56},
  year = {2009},
  doi = {10.1016/j.trb.2008.06.010}
}

@article{AlHaideriWeissIsmailRoundaboutUnderReview,
  title = {Modelling Pedestrians Behaviour at Roundabouts Crosswalks: Exploring a Microscopic Behavioural-Based Approach},
  author = {Al-Haideri, Rulla and Weiss, Adam and Ismail, Karim},
  year = {2026},
  doi = {10.13140/RG.2.2.30619.12329}
}

\end{document}